%% file: main.tex
\documentclass[format=sigconf,10pt]{acmart}
\usepackage[english]{babel} 

\newcommand{\hx}[1]{{\color{red} #1}}

\usepackage{cleveref}
\crefname{section}{}{\S\S}
\crefdefaultlabelformat{\S#2#1#3}

\usepackage{xspace}
\usepackage{comment}
\usepackage[normalem]{ulem}
\usepackage{enumitem}
\setlist{nolistsep}

\usepackage{listings}
\usepackage{color}
\usepackage[font=footnotesize]{caption}

\definecolor{dkgreen}{rgb}{0,0.6,0}
\definecolor{gray}{rgb}{0.5,0.5,0.5}
\definecolor{mauve}{rgb}{0.58,0,0.82}
\newcommand{\sys}{NetKernel\xspace}

\newcommand{\guestlib}{GuestLib\xspace}
\newcommand{\guestlibcore}{Guestlib\_core\xspace}
\newcommand{\guestlibdriver}{GuestLib\_driver\xspace}

\newcommand{\glib}{\guestlib}
\newcommand{\glibcore}{\guestlibcore}

\newcommand{\servicelib}{ServiceLib\xspace}
\newcommand{\servicelibcore}{Servicelib\_core\xspace}
\newcommand{\servicelibdriver}{ServiceLib\_driver\xspace}

\newcommand{\slib}{\servicelib}
\newcommand{\slibcore}{\servicelibcore}
\newcommand{\slibdriver}{\servicelibdriver}

\newcommand{\driver}{nk\_driver\xspace}

\newcommand{\controller}{CoreEngine\xspace}
\newcommand{\nqes}{{NQE}s\xspace}
\newcommand{\nqe}{{NQE}\xspace}

\graphicspath{{figures/}}

\renewcommand\footnotetextcopyrightpermission[1]{} 
\setcopyright{none}

\begin{document}


\title{\sys: Making Network Stack Part of \\the Virtualized
Infrastructure}



\author{Zhixiong Niu}
\affiliation{%
  \institution{City University of Hong Kong}
}

\author{Hong Xu}
\affiliation{%
  \institution{City University of Hong Kong}
}

\author{Peng Cheng}
\affiliation{%
  \institution{Microsoft Research }
}

\author{Yongqiang Xiong}
\affiliation{%
  \institution{Microsoft Research }
}

\author{Tao Wang}
\affiliation{%
  \institution{City University of Hong Kong}
}

\author{Dongsu Han}
\affiliation{%
  \institution{KAIST}
}

\author{Keith Winstein}
\affiliation{%
  \institution{Stanford University}
}

\renewcommand{\shortauthors}{Z. Niu et al.}

%
%



\begin{abstract}

The network stack is implemented inside virtual machines (VMs) in today's
cloud. This paper presents a system called \sys that decouples the network
stack from the guest, and offers it as an independent module implemented by
the cloud operator. \sys represents a new paradigm where network stack is
managed by the operator as part of the virtualized infrastructure. It provides
important efficiency benefits: By gaining control and visibility of the
network stack, operator can perform network management more directly and
flexibly, such as multiplexing VMs running different applications to the same
network stack module to save CPU cores, and enforcing fair bandwidth sharing
with distributed congestion control. Users also benefit from the simplified
stack deployment and better performance. For example mTCP can be deployed
without API change to support nginx and redis natively, and shared memory
networking can be readily enabled to improve performance of colocating VMs.
Testbed evaluation using 100G NICs shows that \sys preserves the performance
and scalability of both kernel and userspace network stacks, and provides the
same isolation as the current architecture.


\end{abstract}

\acmDOI{}

\acmISBN{}


\acmPrice{}

\maketitle

\input{intro}
\input{motivation}

\input{highlight}
\input{design}
\input{implementation}

\input{usecases}

\input{evaluation}
\input{discussion}

\input{related}
\section{Conclusion}

We have presented \sys, a system that decouples the network stack from the
guest, therefore making it part of the virtualized infrastructure in the
cloud. \sys improves network management efficiency for operator, and provides
deployment and performance gains for users. 
We experimentally demonstrated new use cases enabled by \sys that are
otherwise difficult to realize in the current architecture. 
Through testbed evaluation with 100G NICs, 
we showed that \sys achieves the same performance and isolation as today's
cloud.

\sys opens up new design space with many possibilities.
As future work we are implementing zerocopy to the NSM, and exploring using hardware queues
of a SmartNIC to offload \controller and eliminate CPU overhead as in \cref{sec:overhead}.

This work does not raise any ethical issues.

\clearpage
\bibliographystyle{abbrv}
\bibliography{main}



\end{document}

%% file: intro.tex

\section{Introduction}
\label{sec:intro}

Virtual machine (VM) is the predominant virtualization form in today's cloud
due to its strong isolation guarantees. VMs allow customers to run
applications in a wide variety of operating systems (OSes) and configurations.
VMs are also heavily used by cloud operators to deploy internal services, such
as load balancing, proxy, VPN, etc., both in a public cloud for tenants and in
a private cloud for supporting various business units of an organization.
Lightweight virtualization technologies such as containers are also
provisioned inside VMs in many production settings for isolation, security,
and management reasons \cite{ecs,azure-c,google-c}.

\begin{figure}[t]
    \centering
    \includegraphics[width=0.95\linewidth]{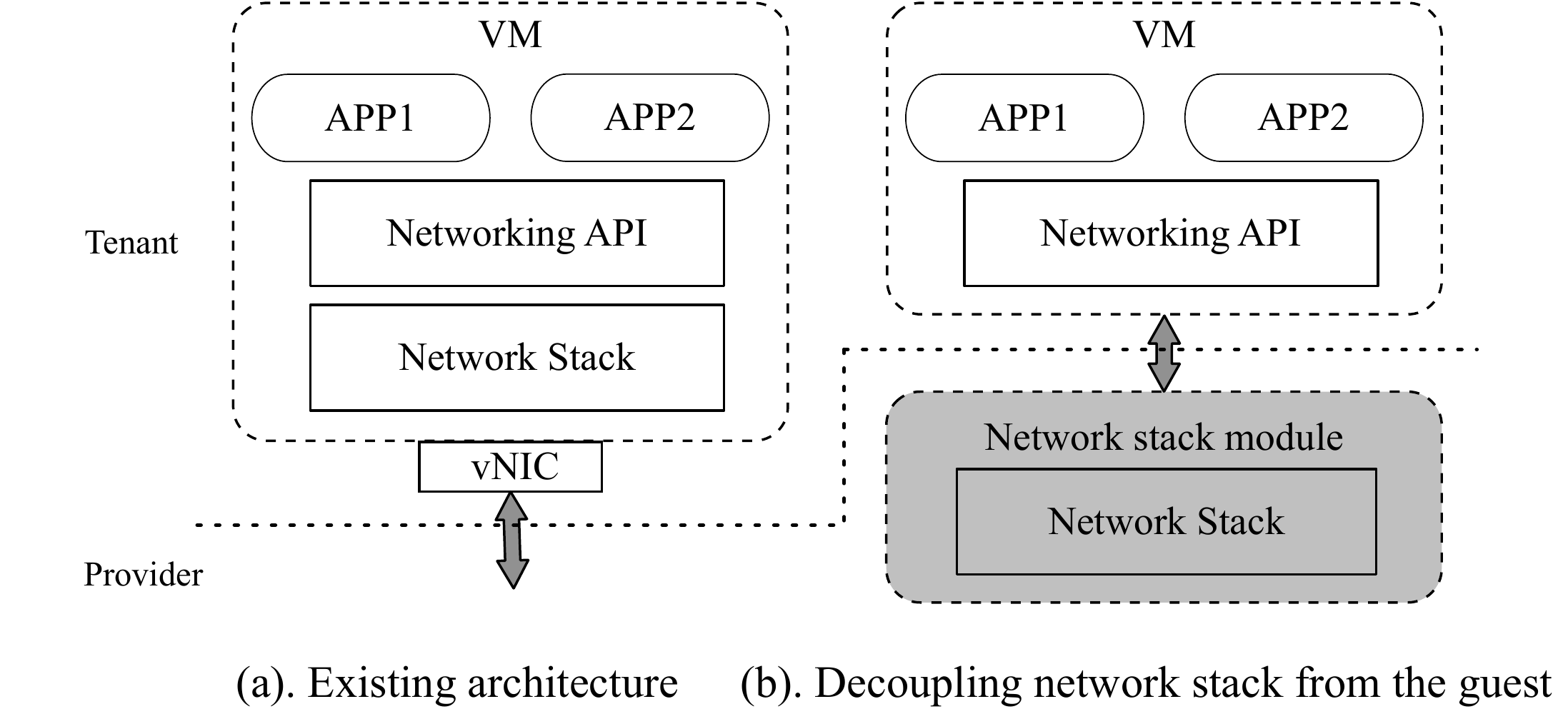}
    \vspace{-2mm}
    \caption{Decoupling network stack from the guest, and making it part of
    the virtualized infrastructure.}
    \label{fig:arch}
\end{figure}

VM based virtualization largely follows traditional OS design. In
particular, the TCP/IP network stack is encapsulated inside the VM as part of the guest OS as shown in
Figure~\ref{fig:arch}(a). Applications own the network stack, which is
separated from the network infrastructure that operators own; they interface
using the virtual NIC abstraction. This architecture preserves the familiar
hardware and OS abstractions so a vast array of workloads can be easily moved
into the cloud. It provides high flexibility to applications to customize
the entire network stack.

We argue that the current division of labor between application and network
infrastructure is becoming increasingly inadequate. The central issue is that
the operator has
almost zero visibility and control over the network stack. This leads to many
efficiency problems that manifest in various aspects of running the cloud
network.

Many network management tasks like monitoring, diagnosis, and
troubleshooting have to be done in an extra layer outside the VMs, which
requires significant effort in design and implementation
\cite{SDVT17,F17,SKGK11}. Since these network functions need to process
packets at the end-host \cite{GYXD15,ZKCG15,KAS19,moshref2016trumpet}, they
can be done more efficiently if the network stack were opened up to
the operator. More importantly, the operator is unable to orchestrate resource
allocation at the end-points of the network fabric, resulting in low resource
utilization. It remains difficult today for the
operator to meet or define performance SLAs despite much prior work
\cite{GLWY10,LMBT14,BCKR11a,PKCK12,JAMP13,PYBM13}, as she cannot precisely
provision resources just for the network stack or control how the stack
consumes these resources. Further, resources (e.g. CPU) have to be provisioned
on a per-VM basis based on the peak traffic; it is impossible to coordinate
across VM boundaries. This degrades the overall utilization of the network
stack since in practice traffic to individual VMs is extremely bursty.

Even the simple task of maintaining or deploying a network stack suffers
from inefficiency today. Network stack has critical impact on performance, and
many optimizations have been studied with numerous effective solutions,
ranging from congestion control \cite{AGMP10,CCGY17,NCRG18}, scalability
\cite{lin2016scalable,JWJJ14}, zerocopy datapath
\cite{YHSE16,JWJJ14,f-stack,ZEFG15,PLZP14}, NIC multiqueue scheduling
\cite{SSAS17}, etc. Yet the operator, with sufficient expertise and resources,
still cannot deploy these extensions to improve performance and reduce
overheads. As a result, our community is still finding ways to deploy DCTCP in
the public cloud \cite{he2016ac,CBVR16}. On the other hand, applications without
much
knowledge of the underlying network or expertise on networking are forced to
juggle the deployment and maintenance details. For example if one wants
to deploy a new stack like mTCP \cite{JWJJ14}, a host of 
problems arise such as setting up kernel bypass, testing with kernel versions
and NIC drivers, and porting applications to the new APIs. Given the intricacy
of implementation and the velocity of development,
it is a daunting task if not impossible to expect users, whether tenants
in a public cloud or first-party services in a private cloud, to individually
maintain the network stack themselves.

We thus advocate a new division of labor in a VM-based cloud in this paper. We
believe that network stack should be managed as part of the
virtualized infrastructure instead of in the VM by application. The operator is
naturally in a better position to own
the last mile of packet delivery, so it can directly deploy, manage,
and optimize the network stack, and comprehensively improve the efficiency of
running the entire network fabric. Applications' functionality and performance
requirements can be consolidated and satisfied with several different network
stacks provided by the operator. As the heavy-lifting is taken care of,
applications can just use network stack as a basic service of the
infrastructure and focus on their business logic.

Specifically, we propose to decouple the VM network stack from the guest as
shown in Figure~\ref{fig:arch}(b). We keep the network APIs such as BSD
sockets intact, and use them as the abstraction boundary between application
and infrastructure. Each VM is served by a network stack module (NSM) that
runs the network stack chosen by the user. Application data are 
handled outside the VM in the NSM, whose design and implementation are managed
by the operator. Various network stacks can be provided as different NSMs to
ensure applications with diverse requirements can be properly satisfied. 
We do not enforce a single transport design, or trade off flexibility of the
existing architecture in our approach.



We make three specific contributions.
\begin{itemize}
\item We design and implement a system called \sys to show that this new
division
of labor is feasible without radical changes to
application or infrastructure (\cref{sec:choices}--\cref{sec:implementation}).
\sys
provides transparent BSD socket redirection so existing applications 
can run directly. The socket semantics from the application are encapsulated
into small queue elements and transmitted to the corresponding NSM via
lockless shared memory queues.

	
\item We present new use cases that are difficult to realize today to show
\sys's potential benefits (\cref{sec:usecases}). For example,  we show
that \sys enables multiplexing: one NSM can serve multiple VMs at the
same time and save over 40\% CPU cores without degrading performance
using traces from a production cloud.

\item We conduct comprehensive testbed evaluation with commodity 100G NICs to
show that \sys achieves the same performance, scalability, and isolation
as the current architecture (\cref{sec:evaluation}). For example, the kernel stack NSM achieves
100G
send throughput with 3 cores; the mTCP NSM achieves 1.1M RPS with 8 cores. 
\end{itemize}

%% file: motivation.tex

\section{Motivation}
\label{sec:motivation}

Decoupling the network stack from the guest OS, hence making it part of the
infrastructure, marks a clear departure from the way networking is
provided to VMs nowadays. In this section we elaborate why this is a better
architectural design by presenting its benefits and contrasting with
alternative solutions. We discuss its potential issues in
\cref{sec:discussion}.



\subsection{Benefits }
\label{sec:benefits}
We highlight key benefits of our vision with new use cases we 
experimentally realize with \sys in \cref{sec:usecases}.

\noindent{\bf Better efficiency in management for the operator.}
Gaining control over the network stack, the operator can now perform network
management more efficiently. For example it can orchestrate the
resource provisioning strategies much more flexibly: For mission-critical
workloads, it can dedicate CPU resources to their NSMs to offer
performance SLAs in terms of throughput and RPS (requests per second)
guarantees. For elastic workloads, on the other hand, it can consolidate
their VMs to the same NSM (if they use the same network stack) to improve
its resource utilization. The operator can also directly implement management
functions as an integral part of user's network stack and improve the
effectiveness of management, compared to doing them in an extra layer
outside the guests.


{\em Use case 1: Multiplexing (\cref{sec:multiplex}).} 
Utilization of network stack in VMs is very low most of the time in practice.
Using a real trace from a large cloud, we show that \sys enables
multiple VMs to be multiplexed onto one
NSM to serve the aggregated traffic and saves over 40\% CPU cores for the
operator without performance
degradation.  

{\em Use case 2: Fair bandwidth sharing (\cref{sec:seawall}).} 
TCP's notion of flow-level fairness
leads to poor bandwidth sharing in data centers \cite{SKGK11}. We show that
\sys allows us to readily implement VM-level congestion control \cite{SKGK11}
as an NSM to achieve fair sharing regardless of
number of flows and destinations.

\noindent{\bf Deployment and performance gains for users.}
Making network stack part of the virtualized infrastructure is also beneficial
for users in both public and private clouds. Operator can directly optimize
the network stack design and implementation. Various kernel stack
optimizations \cite{lin2016scalable,YHSE16}, high-performance userspace stacks
\cite{JWJJ14,Seastar:Website,BPKG14,PLZP14}, and even designs using advanced
hardware \cite{netronome,melalnox,arria,LCLT18} can now be deployed and
maintained transparently without user involvement or application code change. 
Since the BSD socket is the only abstraction exposed to the applications, it
is now feasible to adopt new stack designs independent of the guest kernel or
the network API. 
Our vision also opens up new design space by allowing the network stack to
exploit the visibility into the infrastructure for performance benefits. 

{\em Use case 3: Deploying mTCP without API change (\cref{sec:usecase_mtcp}).} 
We show that \sys 
enables unmodified applications in the VM to use mTCP \cite{JWJJ14} in the
NSM, and improves performance greatly due to mTCP's kernel bypass design. mTCP
is a userspace stack with new APIs (including modified {\tt epoll}/{\tt
kqueue}). During the process, we also find and fix a compatibility issue
between mTCP and our NIC driver, and save significant maintenance time and
effort for users.


{\em Use case 4: Shared memory networking (\cref{sec:sharedmem}).} 
When two VMs of the same user are colocated on the same host, \sys can
directly detect this and copy their data via shared memory to
bypass TCP stack processing and improve throughput.
This is difficult to achieve today as VMs have no knowledge
about the underlying infrastructure \cite{YNRL16,ZZZL19}.

\noindent{\bf And beyond.}
We focus on efficiency benefits in this paper since they seem most immediate.
Making network stack part of the virtualized infrastructure also brings 
additional benefits that are more far-reaching. For example, it facilitates
innovation by allowing new protocols in different layers of the stack
to be rapidly prototyped and experimented. It provides a direct path for
enforcing centralized control, so network functions like failure detection 
\cite{GYXD15} and monitoring \cite{moshref2016trumpet,KAS19} can be integrated
into the network stack implementation. It opens up new design space to
more freely exploit end-point coordination \cite{POBS14,GNKA15},
software-hardware co-design, and programmable
data planes \cite{AGRW17,LCLT18}. 
We encourage the community to fully explore these opportunities in the future.

\subsection{Alternative Solutions }
\label{sec:alternatives}

We now discuss several alternative solutions and why they are inadequate. 

\noindent{\bf Why not just use containers?}
Containers are gaining popularity as a lightweight and portable alternative to
VMs \cite{docket-pop}. A container is essentially a process with namespace
isolation. Using containers can address some efficiency problems because the
network stack is in the hypervisor instead of in the containers. Without the
guest OS, however, containers have poor isolation \cite{KRFX18} and are
difficult to manage. Moreover, containers are constrained to using the host
network stack, whereas \sys provides choices for
applications on the same host. This is important as data center applications
have diverse requirements that cannot be satisfied with a single design.

In a word, containers or other lightweight virtualization represent a more
radical approach of removing the guest kernel, which leads to several
practical issues. Thus they are commonly deployed inside VMs in production
settings. In fact we find that all major public clouds
\cite{ecs,azure-c,google-c} require users to launch containers inside VMs.
Thus, our discussion is centered around VMs that cover the vast majority of
usage scenarios in a cloud. \sys readily benefits
containers running inside VMs as well.

\noindent{\bf Why not on the hypervisor?}
Another possible approach is to keep the VM intact, and add the network stack
implementation outside on the hypervisor. Some existing work takes this
approach to realize a uniform congestion control without changing VMs
\cite{CBVR16,he2016ac}. This does allow the operator to gain control on
network stack. Yet the performance and efficiency of this approach is even
lower than the current architecture because data are then processed twice in
two independent stacks, first by the VM network stack and then the stack
outside.

\noindent{\bf Why not use customized OS images?}
Operators can build customized OS images with the required network stacks for
users, which remedies the maintenance and deployment issues. Yet this approach
has many downsides. It is not transparent: customers need to update these
images on their own, and deploying images causes downtime and disrupts
applications. But more importantly, since even just a single user  
will have vastly different workloads that require different environments
(Linux or FreeBSD or Windows, kernel versions, driver versions, etc.), the
cost of testing and maintenance for all these
possibilities is prohibitive.

In contrast, \sys does not have these issues because it breaks the coupling of
the network stack to the guest OS. Architecturally a network stack module can
be used by VMs with different guest OSes since BSD socket APIs are widely
supported, thereby greatly reducing development resources required for
operators. Maintenance is also transparent and non-disruptive to customers as
operators can roll out updates in the background.

%% file: highlight.tex

\section{Design Philosophy}
\label{sec:choices}

\sys imposes three fundamental design questions around the separation of network
stack and the guest OS: 
\begin{enumerate}
\item How to {transparently} redirect socket API calls without changing applications? 
\item How to transmit the socket semantics between the VM and NSM whose implementation of the stack may vary? 
\item How to ensure high performance with semantics transmission (e.g., 100\,Gbps)?
\end{enumerate}
These questions touch upon largely uncharted territory in the design space.  
Thus our main objective in this paper is to demonstrate feasibility of our approach on existing virtualization platforms and showcase its potential.
Performance and overhead are not our primary goals. 
It is also not our goal to improve any particular network stack design.

In answering the questions above, \sys's design has the following highlights. 

\noindent{\bf Transparent Socket API Redirection.} 
\sys needs to redirect BSD socket calls to the NSM instead of the tenant network stack. 
This is done by inserting into the guest a library called \guestlib. 
The \guestlib provides a new socket type called \sys socket with a complete implementation of BSD socket APIs. 
It replaces all TCP and UDP sockets when they are created with \sys sockets, effectively redirecting them without changing applications.

\noindent{\bf A Lightweight Semantics Channel.} 
Different network stacks may run as different NSMs, so \sys needs to ensure socket semantics from the VM work properly with the actual NSM stack implementation. 
For this purpose \sys builds a lightweight socket semantics channel between VM and its NSM. 
The channel relies on small fix-sized queue elements as intermediate representations of socket semantics:
each socket API call in the VM is encapsulated into a queue element and sent to the NSM, 
who would effectively translate the queue element into the corresponding API call of its network stack.

\noindent{\bf Scalable Lockless Queues.} 
As NIC speed in cloud evolves from 40G/50G to 100G \cite{FPMC18} and higher, the NSM has to use multiple cores for the network stack to achieve line rate.
\sys thus adopts scalable lockless queues to ensure VM-NSM socket semantics transmission is not a bottleneck. 
Each core services a dedicated set of queues so performance is scalable with number of cores. 
More importantly, each queue is memory shared with a software switch, so it can be lockless with only a single producer and a single consumer to avoid expensive lock contention \cite{HRW14,JWJJ14,lin2016scalable}.

Switching the queue elements offers important benefits beyond lockless queues.
It facilitates a flexible mapping between VM and NSM: a NSM can support
multiple VMs without adding more queues compared to binding the queues
directly between VM and NSM. In addition, it allows dynamic resource
management: cores can be readily added to or removed from a NSM, and a user
can switch her NSM on the fly. 
The CPU overhead of software switching can be addressed by hardware offloading \cite{FPMC18,sigcomm15keynote}, which we discuss in \cref{sec:overhead} in more detail.

\noindent{\bf VM Based NSM.}
Lastly we discuss an important design choice regarding the NSM.  
The NSM can take various forms. It may be a full-fledged VM with a monolithic
kernel.  
Or it can be a container or module running on the hypervisor, which is
appealing because it consumes less resource and offers better performance. Yet
it entails porting a complete TCP/IP stack to the hypervisor. Achieving memory
isolation among containers or modules are also difficult \cite{PHJW16}. More
importantly, it introduces another coupling between the network stack and the
hypervisor, which defeats the purpose of \sys. Thus we choose to use a VM
for NSM. VM based NSM readily supports existing kernel and userspace stacks
from various OSes. VMs also provide good isolation and we can dedicate
resources to a NSM to guarantee performance. VM based NSM is the most
flexible: we can run stacks independent of the hypervisor.

%% file: design.tex
\section{Design}
\label{sec:design}

\begin{figure}[t]
    \centering
    \includegraphics[width=1\linewidth]{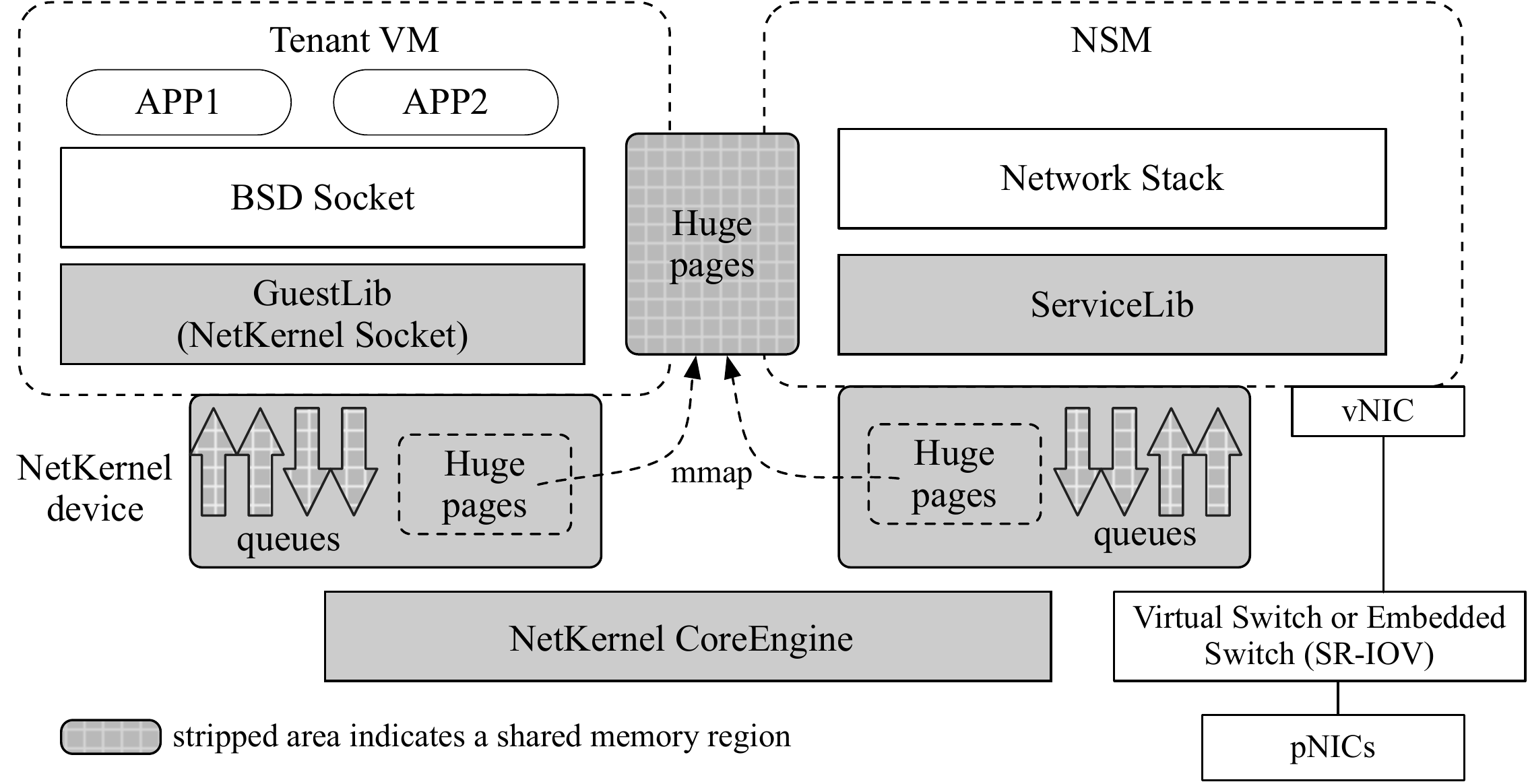}
    \caption{NetKernel design overview. }
    \label{fig:design_VMs}
    \vspace{-3mm}
\end{figure}

Figure~\ref{fig:design_VMs} depicts NetKernel's architecture. 
The BSD socket APIs are transparently redirected to a complete NetKernel socket implementation in \guestlib in the guest kernel  (\cref{sec:api_redirection}). 
The \guestlib can be deployed as a kernel patch and is the only change we make to the user VM.  
Network stacks are implemented by the provider on the same host as Network Stack Modules (NSMs), which are individual VMs in our current design. 
Inside the NSM, a \servicelib interfaces with the network stack. 
The NSM connects to the vSwitch, be it a software or a hardware switch, and then the pNICs. Thus our design also supports SR-IOV. 

All socket operations and their results are translated into NetKernel Queue Elements (\nqes) by \guestlib and \servicelib (\cref{sec:semantics_channel}).
For \nqe transmission, \guestlib and \servicelib each has a NetKernel device, or NK device in the following, consisting of one or more sets of lockless queues. 
Each queue set has a {{\em send queue} and {\em receive queue} for operations with data transfer (e.g. {\tt send()}), and a {\em job queue} and {\em completion queue} for control operations without data transfer (e.g. {\tt setsockopt()})}. 
Each NK device connects to a software switch called \controller, which runs on the hypervisor and performs actual \nqe switching (\cref{sec:nqe_switching}).
The \controller is also responsible for various management tasks such as setting up the NK devices, ensuring isolation among VMs, etc. (\cref{sec:ce_control}) 
A unique set of hugepages are shared between each VM-NSM tuple for application data exchange.
A NK device also maintains a hugepage region that is memory mapped to the corresponding application hugepages as shown in Figure~\ref{fig:design_VMs} (\cref{sec:data}). 

For ease of presentation, we assume both the user VM and NSM run Linux, and the
NSM uses the kernel stack.

\subsection{\bf Transparent Socket API Redirection}
\label{sec:api_redirection}

We first describe how NetKernel's \guestlib interacts with applications to support BSD socket semantics transparently. 

\noindent{\bf Kernel Space API Redirection.} 
There are essentially two approaches to redirect BSD socket calls to NSM, each with its unique tradeoffs. 
One is to implement it in userspace using {\tt LD\_PRELOAD} for example. 
The advantages are: (1) It is efficient without syscall overheads and performance is high \cite{JWJJ14}; (2) It is easy to deploy without kernel modification.
However, this implies each application needs to have its own redirection service, which limits the usage scenarios. 
Another way is kernel space redirection, which naturally supports multiple applications without IPC. 
The flip side is that performance may be lower due to context switching and syscall overheads.

We opt for kernel space API redirection to support most of the usage scenarios, and leave userspace redirection as future work.
\guestlib is a kernel module deployed in the guest. 
This is feasible by distributing images of para-virtualizated guest kernels to users, a practice providers are already doing nowadays. 
Kernel space redirection also allows NetKernel to work directly with I/O event notification syscalls like {epoll}.

\noindent{\bf NetKernel Socket API.}
\guestlib creates a new type of sockets---{\tt SOCK\_NETKERNEL}, in addition to TCP ({\tt SOCK\_STREAM}) and UDP ({\tt SOCK\_DGRAM}) sockets. 
It registers a complete implementation of BSD socket APIs as shown in Table~\ref{table:design_netkernelsocket} to the guest kernel. 
When the guest kernel receives a {\tt socket()} call to create a new TCP socket say, it replaces the socket type with {\tt SOCK\_NETKERNEL}, creates a new NetKernel socket, and initializes the socket data structure with function pointers to NetKernel socket implementation in \guestlib. 
The {\tt sendmsg()} for example now points to {\tt nk\_sendmsg()} in \guestlib instead of {\tt tcp\_sendmsg()}.
\begin{table}[h]
\centering
\caption{NetKernel socket implementation.}
\label{table:design_netkernelsocket}
\resizebox{\columnwidth}{!}{%
\begin{tabular}{|l|l|l|}
\hline
           & inet\_stream\_ops        & netkernel\_pro       \\ \hline
bind       & {\tt inet\_bind()}               & {\tt nk\_bind()}       \\ \hline
connect    & {\tt inet\_connect()}	        & {\tt nk\_connect()}    \\ \hline
accept     & {\tt inet\_accept()}             & {\tt nk\_accept()}     \\ \hline
poll       & {\tt tcp\_poll()}                & {\tt nk\_poll()}       \\ \hline
ioctl      & {\tt inet\_ioctl()}              & {\tt nk\_ioctl()}      \\ \hline
listen     & {\tt inet\_listen()}             & {\tt nk\_listen()}     \\ \hline
shutdown   & {\tt inet\_shutdown()}           & {\tt nk\_shutdown()}   \\ \hline
setsockopt & {\tt sock\_common\_setsockopt()} & {\tt nk\_setsockopt()} \\ \hline
recvmsg    & {\tt tcp\_recvmsg()}             & {\tt nk\_recvmsg()}    \\ \hline
sendmsg    & {\tt tcp\_sendmsg()}             & {\tt nk\_sendmsg()}    \\ \hline
\end{tabular}
}
\end{table}

\subsection{\bf A Lightweight Socket Semantics Channel}
\label{sec:semantics_channel}

Socket semantics are contained in \nqes and carried around between \guestlib and \servicelib via their respective NK devices.
\begin{figure}[ht]
    \centering
    \includegraphics[width=0.99\linewidth]{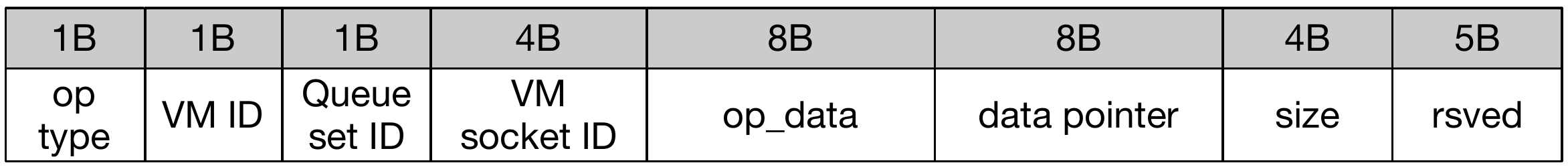}
    \caption{Structure of a \nqe. Here socket ID denotes a pointer to the {\tt sock} struct in the user VM or NSM, and is used for \nqe transmission with VM ID and queue set ID in \cref{sec:nqe_switching}; {\tt op\_data} contains data necessary for socket operations, such as ip address for bind; {\tt data pointer} is a pointer to application data in hugepages; and {\tt size} is the size of pointed data in hugepages.}
    \label{fig:nqe}
\end{figure}

\begin{figure}[ht]
    \centering
    \includegraphics[width=0.65\linewidth]{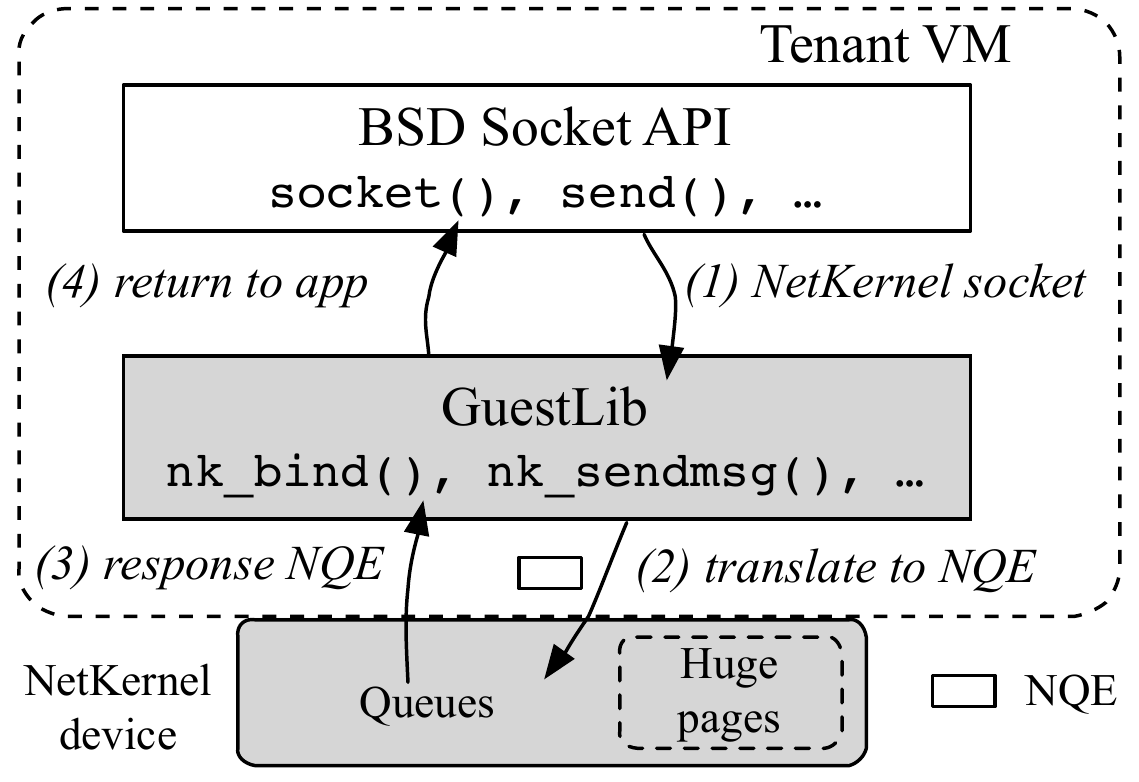}
    \caption{NetKernel socket implementation in \guestlib redirects socket API calls. \guestlib translates  socket API calls to \nqes and \servicelib translates results into \nqes as well (not shown here).}
    \label{fig:design_guestlib}
\end{figure}

\noindent{\bf \nqe and Socket Semantics Translation.}
Figure~\ref{fig:nqe} shows the structure of a \nqe with a fixed size of 32 bytes. 
Translation happens at both ends of the semantics channel: 
\guestlib encapsulates the socket semantics into \nqes and sends to \servicelib, which then invokes the corresponding API of its network stack to execute the operation; 
the execution result is again turned into a \nqe in \servicelib first, and then translated by \guestlib back into the corresponding response of socket APIs.

For example in Figure~\ref{fig:design_guestlib}, to handle the {\tt socket()} call in the VM, \guestlib creates a new \nqe with the operation type and information such as its VM ID for \nqe transmission. 
The \nqe is transmitted by \guestlib's NK device.
The {\tt socket()} call now blocks until a response \nqe is received.
After receiving the \nqe, \servicelib parses the \nqe from its NK device, invokes the {\tt socket()} of the kernel stack to create a new TCP socket, prepares a new \nqe with the execution result, and enqueues it to the NK device. 
\guestlib then receives and parses the response \nqe and wakes up the {\tt socket()} call.
The {\tt socket()} call now returns to application with the NetKernel socket file descriptor (fd) if a TCP socket is created at the NSM, or with an error number consistent with the execution result of the NSM. 

We defer the handling of application data to \cref{sec:data}.


\noindent{\bf Queues for \nqe Transmission.}
\nqes are transmitted via one or more sets of queues in the NK devices.
A queue set has four independent queues: 
a {\em job queue} for \nqes representing socket operations issued by the VM without data transfer,  
a {\em completion queue} for \nqes with execution results of control operations from the NSM, 
a {\em send queue} for \nqes representing operations issued by VM with data transfer; 
and a {\em receive queue} for \nqes representing events of newly received data from NSM. 
Queues of different NK devices have strict correspondence: 
the \nqe for {\tt socket()} for example is put in the job queue of \guestlib's NK device, and sent to the job queue of \servicelib's NK device.

We now present the working of I/O event notification mechanisms like epoll with the receive queue. 
Suppose an application issues {\tt epoll\_wait()} to monitor some sockets. 
Since all sockets are now NetKernel sockets, the {\tt nk\_poll()} is invoked by {\tt epoll\_wait()} and checks the receive queue to see if there is any \nqe for this socket. 
If yes, this means there are new data received, {\tt epoll\_wait()} then returns and the application issues a {\tt recv()} call with the NetKernel socket fd of the event. This points to {\tt nk\_recvmsg()} which parses the \nqe from receive queue for the data pointer, copies data from the hugepage directly to the userspace, and returns. 

If {\tt nk\_poll()} does not find any relevant \nqe, it sleeps until \controller wakes up the NK device when new \nqes arrive to its receive queue.  
\guestlib then parses the \nqes to check if any sockets are in the epoll instances, and wakes up the epoll to return to application. 
{An {\tt epoll\_wait()}can also be returned by a timeout}.

    \begin{figure}[ht]
    \centering
    \includegraphics[width=1\linewidth]{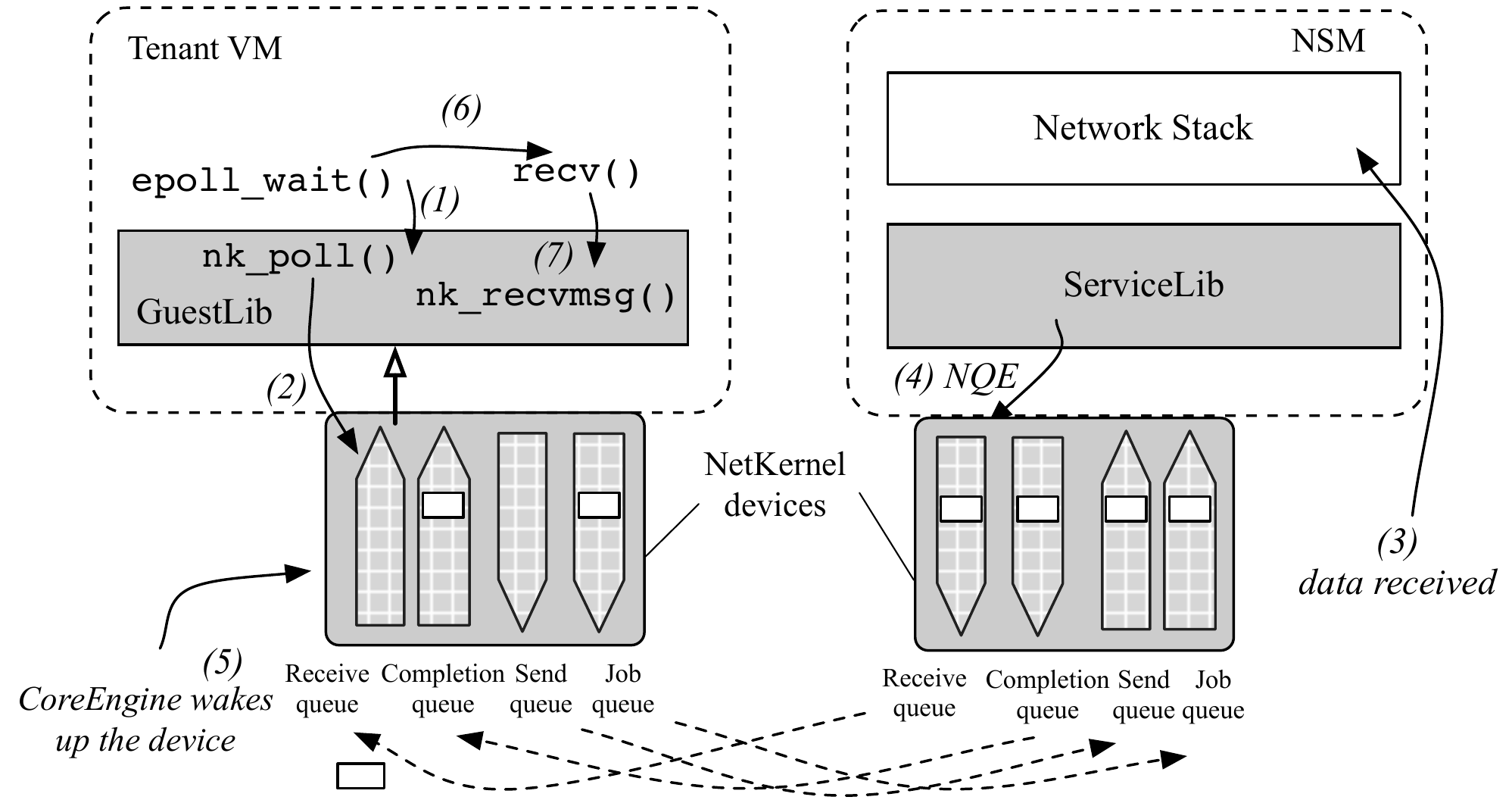}
    \caption{The socket semantics channel with epoll as an example. \guestlib and \servicelib translate semantics to \nqes, and queues in the NK devices perform \nqe transmission. Job and completion queues are for socket operations and execution results, send queues are for socket operations with data, and receive queues are for events of newly received data. Application data processing is not shown. }
    \label{fig:epoll}
\end{figure}

\subsection{\nqe Switching across Lockless Queues}
\label{sec:nqe_switching}

We now elaborate how \nqes are switched by \controller and how the NK devices interact with \controller.

\noindent{\bf Scalable Queue Design.}
The queues in a NK device is scalable: there are one dedicated queue set per vCPU for both VM and NSM, so NetKernel performance scales with CPU resources. 
Each queue set is shared memory with the \controller, essentially making it a single producer single consumer queue without lock contention. 
VM and NSM may have different numbers of queue sets. 

\begin{figure}[ht]
    \centering
    \includegraphics[width=1\linewidth]{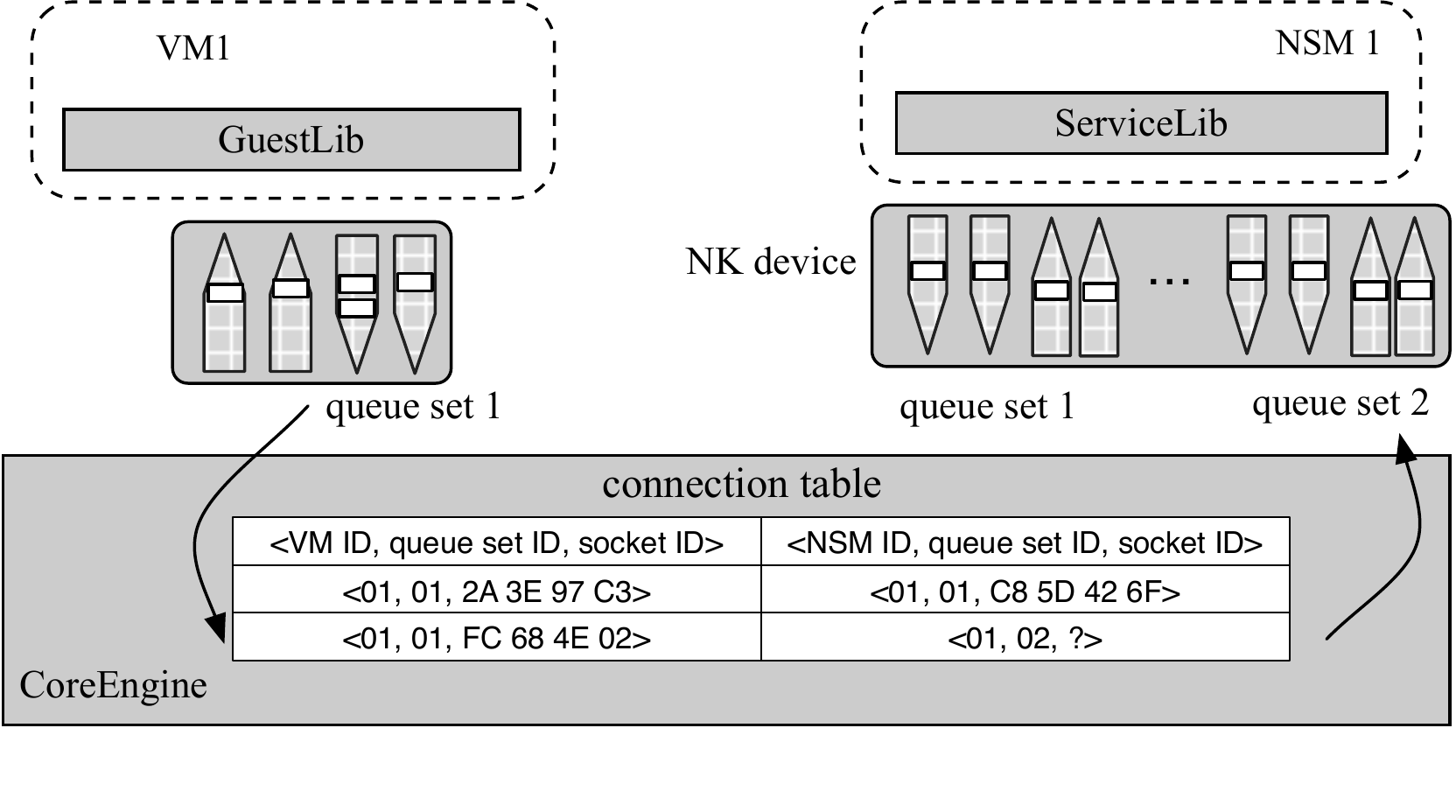}
    \caption{\nqe switching with \controller. }
    \label{fig:nqe_trans}
\end{figure}

\noindent{\bf Switching \nqes in \controller.}
\nqes are load balanced across multiple queue sets  with the \controller acting as a switch. 
\controller maintains a connection table as shown in Figure~\ref{fig:nqe_trans}, which maps the tuple $\langle$VM ID, queue set ID, socket ID$\rangle$ to the corresponding $\langle$NSM ID, queue set ID, socket ID$\rangle$ and vice versa. 
Here a socket ID corresponds to a pointer to the {\tt sock} struct in the user VM or NSM. 
We call them VM tuple and NSM tuple respectively. 
\nqes only contain VM tuple information.

Using the running example of the {\tt socket()} call, we can see how \controller uses the connection table. The process is also shown in Figure~\ref{fig:nqe_trans}. 
(1) When \controller processes the socket \nqe from VM1's queue set 1, it realizes this is a new connection, and 
inserts a new entry to the table with the VM tuple from the \nqe.
(2) It checks which NSM should handle it,\footnote{A user VM to NSM mapping is determined either by the users offline or some load balancing scheme dynamically by \controller.} 
performs hashing based on the three tuple to determine which queue set (say 2) to switch to if there are multiple queue sets, 
and copies the \nqe to the NSM's corresponding job queue. 
\controller adds the NSM ID and queue set ID to the new entry. 
(3) \servicelib gets the \nqe and copies the VM tuple to its response \nqe, and adds the newly created connection ID in the NSM to the {\tt op\_data} field of response \nqe. 
(4) \controller parses the response \nqe, matches the VM tuple to the entry and adds the NSM socket ID to complete it, and copies the response \nqe to the completion queue 1 of the corresponding VM1 as instructed in the \nqe. 
Later \nqes for this VM connection can be processed by the correct NSM connection and vice versa. 
Note \servicelib pins its connections to its vCPUs and queue sets, thus processing the \nqe and sending the response \nqe is done on the same CPU. 

The connection table allows flexible multiplexing and demultiplexing with the socket ID information. For example one NSM can serve multiple VMs using different sockets. 
{\controller uses polling across all queue sets to maximize performance.}

\subsection{Management with \controller}
\label{sec:ce_control}

\controller acts as the control plane of NetKernel and carries out many control tasks beyond \nqe switching.

\noindent{\bf NK Device and Queue Setup.}
\controller allocates shared memory for the queue sets and setts up the 
NK devices accordingly when a VM or NSM starts up, and de-allocates when they shut down.
Queues can also be dynamically added or removed with the number of vCPUs.

\noindent{\bf Isolation.}
\controller sits in an ideal position to carry out isolation among VMs, 
a task essential in public clouds with VMs sharing one NSM.
In our design \controller polls each queue set in a round-robin fashion to ensure the basic fair sharing. 
Providers can implement other forms of isolation mechanisms to rate limit a VM in terms of bandwidth or the number of \nqes (i.e. operations) per second, which we also experimentally show in \cref{sec:isolation_perf}.
Note that \controller isolation happens for egress; ingress isolation at the NSM in general is more challenging and may need to resort to physical NIC queues \cite{DSAA18}.


\subsection{Processing Application Data}
\label{sec:data}

So far we have covered API redirection, socket semantics transmission, \nqe switching, and \controller management in NetKernel.
We now discuss the last missing piece: 
how application data are actually processed in the system.

\noindent{\bf Sending Data.}
Application data are transmitted by hugepages shared between the VM and NSM.
Their NK devices maintain a hugepage region that is mmaped to the application hugepages. 
For sending data with {\tt send()}, \guestlib copies data from userspace directly to the hugepage, and adds a data pointer to the send \nqe.
It also increases the send buffer usage for this socket similar to the send buffer size maintained in an OS. 
The {\tt send()} now returns to the application.
\servicelib invokes the {\tt tcp\_sendmsg()} provided by the kernel stack upon receiving the send \nqe. 
Data are obtained from the hugepage, processed by the network stack and sent via the vNIC.
A new \nqe is generated with the result of send at NSM and sent to \guestlib,
who then decreases the send buffer usage accordingly. 

\noindent{\bf Receiving Data.}
Now for receiving packets in the NSM, a normal network stack would send received data to userspace applications.
In order to send received data to the user VM,  
\servicelib then copies the data chunk to huge pages and create a new \nqe to the receive queue, which is then sent to the VM. 
It also increases the receive buffer usage for this connection, similar to the send buffer maintained by \guestlib described above.
The rest of the receive process is already explained in \cref{sec:semantics_channel}.
Note that application uses {\tt recv()} to copy data from hugepages to their own buffer.

\noindent{\bf \servicelib.}
As discussed \servicelib deals with much of data processing at the NSM side so the network stack works in concert with the rest of NetKernel. 
One thing to note is that unlike the kernel space \guestlib, \servicelib should live in the same space as the network stack to ensure best performance. 
We have focused on a Linux kernel stack with a kernel space \servicelib here. 
The design of a userspace \servicelib for a userspace stack is similar in principle. 
We implement both types of stacks as NSMs in \cref{sec:implementation}. 
{\servicelib polls all its queues whenever possible for maximum performance.}

\subsection{Optimization}
\label{sec:opt_in_nk}

We present several best-practice optimizations employed in NetKernel to improve efficiency.

\noindent{\bf Pipelining.}
NetKernel applies pipelining in general between VM and NSM for performance. 
For example on the VM side, a {\tt send()} returns immediately after putting data to the hugepages, 
instead of waiting for the actual send result from the NSM. 
Similarly the NSM would handle the {\tt accept()} by accepting a new connection and returning immediately, before the corresponding \nqe is sent to \guestlib and then application to process.
Doing so does not break BSD socket semantics. 
Take {\tt send()} for example. A successful {\tt send()} does not guarantee delivery of the message \cite{posix}; it merely indicates the message is written to socket buffer successfully.
In NetKernel a successful {\tt send()} indicates the message is written to buffer in the hugepages successfully.
{As explained in \cref{sec:data} the NSM sends the result of send back to the VM to indicate if the socket buffer usage can be decreased or not. }

\noindent{\bf Interrupt-Driven Polling.}
We adopt an interrupt-driven polling design for \nqe event notification to \guestlib's NK device.
This is to reduce the overhead of \guestlib and user VM.  
When an application is waiting for events e.g. the result of the {\tt socket()} call or receive data for epoll, 
the device will first poll its completion queue and receive queue. 
If no new \nqe comes after a short time period (20\textmu s in our experiments), the device sends an interrupt to \controller, notifying that it is expecting \nqe, and stops polling. 
\controller later wakes up the device, which goes back to polling mode to process new \nqes from the completion queue. 

Interrupt-driven polling presents a favorable trade-off between overhead and performance compared to pure polling based or interrupt based design. 
It saves precious CPU cycles when load is low and ensures the overhead of NetKernel is very small to the user VM. 
Performance on the other hand is competent since the response \nqe is received within the polling period in most cases for blocking calls, and when the load is high polling automatically drives the notification mechanism. %
As explained before \controller and \servicelib both use busy polling to maximize performance.


\noindent{\bf Batching.}
As a common best-practice, batching is used in many parts of NetKernel for better throughput. 
\controller uses batching whenever possible for polling from and copying into the queues. 
The NK devices also receive \nqes in a batch for both \guestlib and \servicelib.

%% file: implementation.tex
\section{Implementation}
\label{sec:implementation}

Our implementation is based on QEMU KVM 2.5.0 and Linux kernel 4.9 for both the host and the guest OSes,
with over 11K LoC.
We plan to open source our implementation.

\noindent{\bf \guestlib.}
We add the {\tt SOCK\_NETKERNEL} socket to the kernel ({\tt net.h}), and modify {\tt socket.c} to rewrite the {\tt SOCK\_STREAM} to {\tt SOCK\_NETKERNEL} during the socket creation.
We implement \guestlib as a kernel module with two components: \glibcore and \driver. 
\glibcore is mainly for Netkernel sockets and \nqe translation, and \driver is for \nqe communications via queues. 
\glibcore and \driver communicate with each other using function calls.

\noindent{\bf \servicelib and NSM.}
We also implement \slib as two components: \slibcore and \driver. 
\slibcore translates \nqes to network stack APIs, and the \driver is identical with the one in \glib. 
For the kernel stack NSM, \slibcore calls the kernel APIs directly to handle socket operations without entering userspace. 
We create an independent {\tt kthread} to poll the job queue and send queue for \nqes to avoid kernel stuck.
Some BSD socket APIs can not be invoked in kernel space directly. 
We use {\tt EXPORT\_SYMBOLS} to export the functions for \slib. 
Meanwhile, the boundary check between kernel space and userspace is disabled. 
We use per-core {\tt epoll\_wait()} to obtain incoming events from the kernel stack.

We also port mTCP \cite{mtcp-version} as a userspace stack NSM. 
It uses DPDK 17.08 as the packet I/O engine.
The DPDK driver has not been tested for 100G NICs 
and we fixed a compatibility bug during the process; more details are in \cref{sec:usecase_mtcp}. 
For simplicity, we maintain the two-thread model and per-core data structure in mTCP. 
We implement the NSM in mTCP's application thread at each core. 
The per-core mTCP thread (1) translates \nqes polled from the NK device to mTCP socket APIs, 
and (2) responds \nqes to the tenant VM based on the network events collected by {\tt mtcp\_epoll\_wait()}. 
Since mTCP works in non-blocking mode for performance enhancement, 
we buffer {send} operations at each core and set the {\tt timeout} parameter to 1ms in {\tt mtcp\_epoll\_wait()} to avoid starvation when polling \nqe requests.



\noindent{\bf Queues and Huge Pages.}
The huge pages are implemented based on QEMU's {IVSHMEM}. The page size is 2~MB and we use 128 pages. The queues are ring buffers implemented as much smaller {IVSHMEM} devices.
Together they form a NK device which is a virtual device to the VM and NSM.



\noindent{\bf \controller.}
{The \controller is a daemon with two threads on the KVM hypervisor. 
One thread listens on a pre-defined port to handle NK device (de)allocation requests, namely 8-byte network messages of the tuples $\langle  ce\_op, ce\_data \rangle$. When a VM (or NSM) starts (or terminates), 
it sends a request to \controller for registering (or deregistering) a NK device. 
If the request is successfully handled, CoreEngine responds in the same message format. 
Otherwise, an error code is returned.}
{The other thread polls \nqes in batches from all NK devices and switches them as described in \cref{sec:nqe_switching}.}


%% file: usecases.tex
\section{New Use Cases}
\label{sec:usecases}

To demonstrate the potential of \sys, we present some new use cases that are realized in our implementation.
Details of our testbed is presented in \cref{sec:setup}.
The first two use cases show benefits for the operator, while the next two show
benefits for users.

\subsection{Multiplexing}
\label{sec:multiplex}

Here we describe a new use case where the operator can optimize
resource utilization by serving multiple bursty VMs
with one NSM.  

\begin{figure}[ht]
    \centering
    \begin{minipage}{.48\linewidth}
    \includegraphics[width=\linewidth]{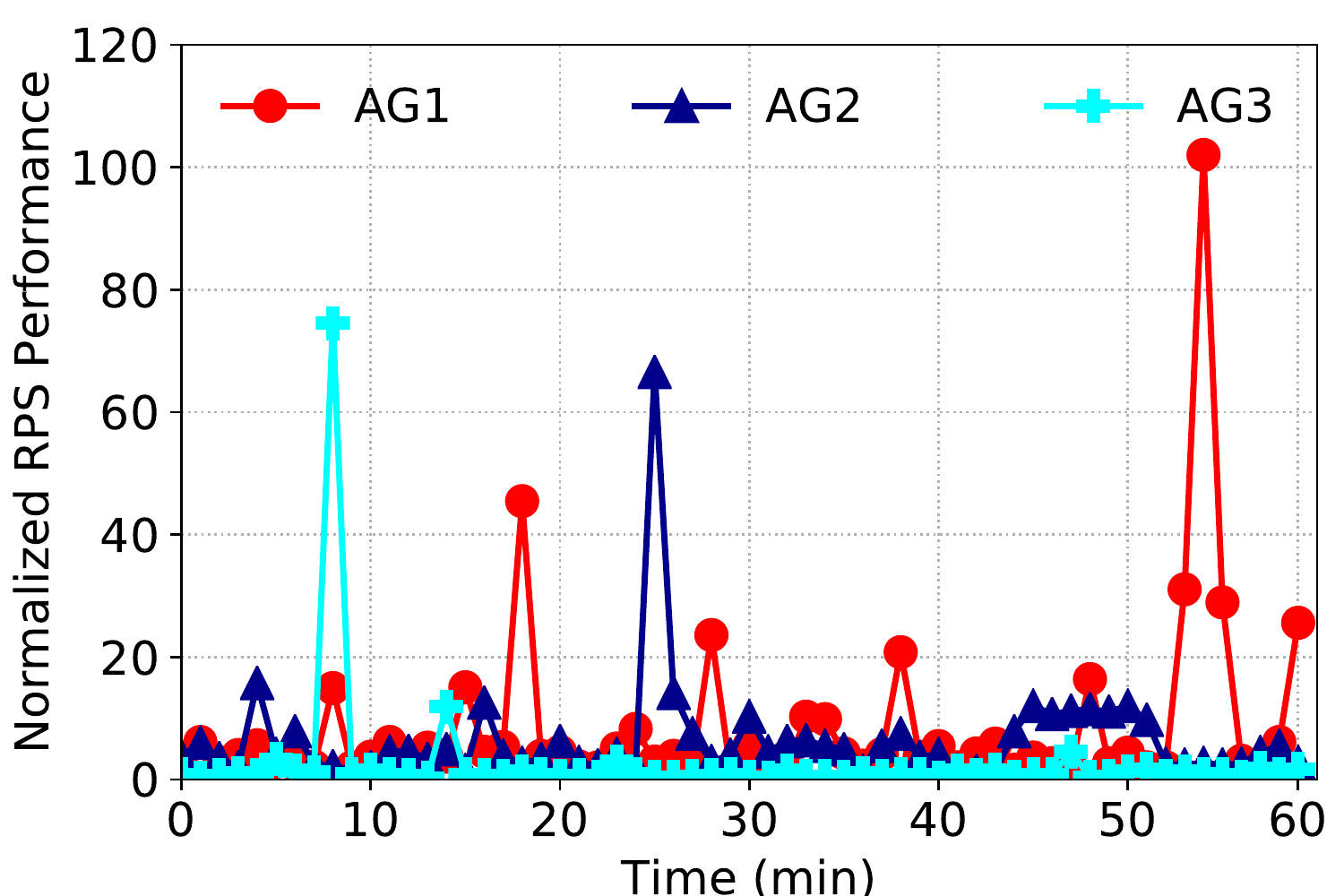}
    \caption{Traffic of three most utilized application gateways (AGs) in our
    trace. They are deployed as VMs. }
    \label{fig:result_multiplex_ms_trace}
    \end{minipage}
    \hspace{0.1cm}
    \begin{minipage}{.48\linewidth}
    \centering
    \includegraphics[width=\linewidth]{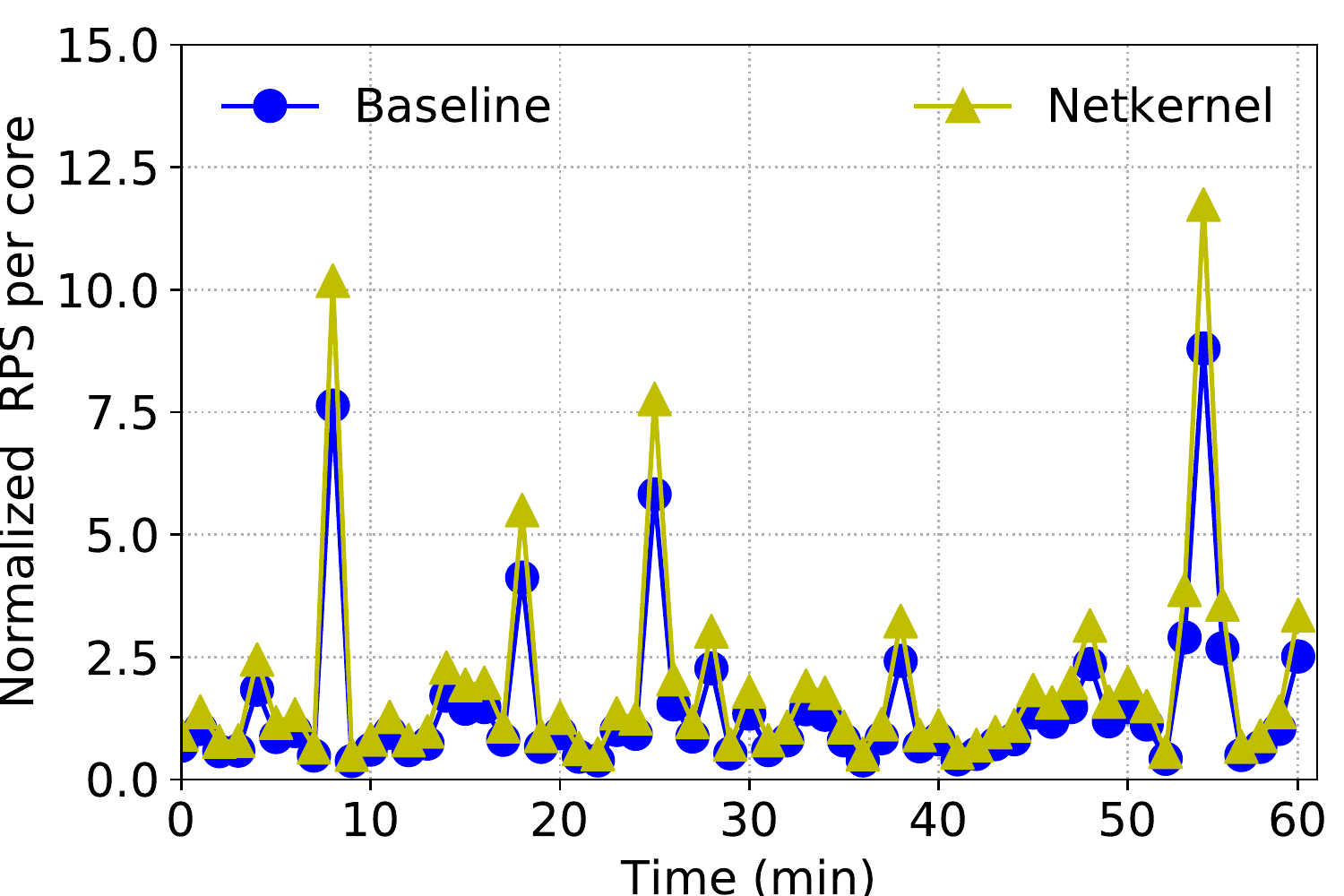}
    \caption{Per-core RPS comparison. Baseline uses 12 cores for 3 AGs, while
    \sys with multiplexing only needs 9 cores. }
    \label{fig:result_multiplex_ms_trace2}
    \end{minipage}
\end{figure}

To make things concrete we draw upon a user traffic trace collected from a
large cloud in September 2018. The trace contains statistics of tens of
thousands of 
application gateways (AGs) that handle tenant (web) traffic in order to
provide load balancing, proxy, and other services. The AGs are internally
deployed as VMs by the operator. We find that the AG's average utilization is
very low most of the time. Figure~\ref{fig:result_multiplex_ms_trace} shows
normalized traffic processed by three most utilized AGs (in the same
datacenter) in our trace with 1-minute intervals for a 1-hour period. We can
clearly see the bursty nature of the traffic. Yet it is very difficult to
consolidate their workloads in current cloud because they serve different
customers using different configurations (proxy settings, LB strategies,
etc.), and there is no way to separate the application logic with the
underlying network stack. The operator has to deploy AGs as independent VMs, 
reserve resources for them, and charge customers accordingly.

\sys enables multiplexing across AGs running distinct services, since the
common TCP stack processing is now separated into the NSM. Using the three
most utilized AGs which have the least benefit from multiplexing as an
example, without \sys each needs 4 cores in our testbed to handle their
peak traffic, and the total per-core requests per second (RPS) of the system
is depicted in Figure~\ref{fig:result_multiplex_ms_trace2} as Baseline. Then
in \sys, we deploy 3 VMs each with 1 core to replay the trace as the AGs,
and use a kernel stack NSM with 5 cores which is sufficient to handle the
aggregate traffic. Totally 9 cores are used including \controller,
representing a saving of 3 cores in this case. The per core RPS is thus
improved by 33\% as shown in Figure~\ref{fig:result_multiplex_ms_trace2}.
Each AG has exactly the same RPS performance without any packet loss. 

In the general case multiplexing these AGs brings even more gains since their
peak traffic is far from their capacity. For ease of exposition we assume the
operator reserves 2 cores for each AG. A 32-core machine can host 16 AGs. If
we use \sys with 1 core for \controller and a 2-core NSM, we find that we
can always pack 29 AGs each with 1 core for the application logic as depicted
in Table~\ref{table:nk_hypervisor}, and the maximum utilization of the NSM
would be well under 60\% in the worst case for $\sim$97\% of the AGs in the
trace. Thus one machine can run 13 or 81.25\% more AGs now, which means the
operator can save over 40\% cores for supporting this workload. 
This implies salient financial gains for the operator: according to 
\cite{FPMC18} one physical core has a maximum potential revenue of \$900/yr. 

\begin{table}[ht]
\centering
\small
\resizebox{0.65\columnwidth}{!}{
    \begin{tabular}{|l|l|l|}
    \hline
                      & Baseline & \sys \\ \hline
    Total \# Cores       & 32           & 32                       \\ \hline
    NSM  & 0            & 2                        \\ \hline
    CoreEngine & 0            & 1                        \\ \hline
    \# AGs         & 16           & 29                       \\ \hline
    \end{tabular}
    }
    \caption{\sys multiplexes more AGs per machine and saves over 40\%
    cores.}
    \label{table:nk_hypervisor}
    \end{table}


\subsection{Fair Bandwidth Sharing}
\label{sec:seawall}
TCP is designed to achieve flow-level fairness for bandwidth sharing in a
network. This leads to poor fairness in a cloud where a misbehaved VM can hog
the bandwidth by say using many TCP flows. Distributed congestion control at
an entity-level (VM, process, etc.) such as Seawall \cite{SKGK11} has been
proposed and implemented in a non-virtualized setting. Yet using Seawall in a
cloud has many difficulties: the provider has to implement it on the vSwitch
or hypervisor and make it work for various guest OSes. The interaction with
the VM's own congestion control logic makes it even harder \cite{he2016ac}.

\begin{figure}[ht]
    \centering
    \includegraphics[width=0.8\linewidth]{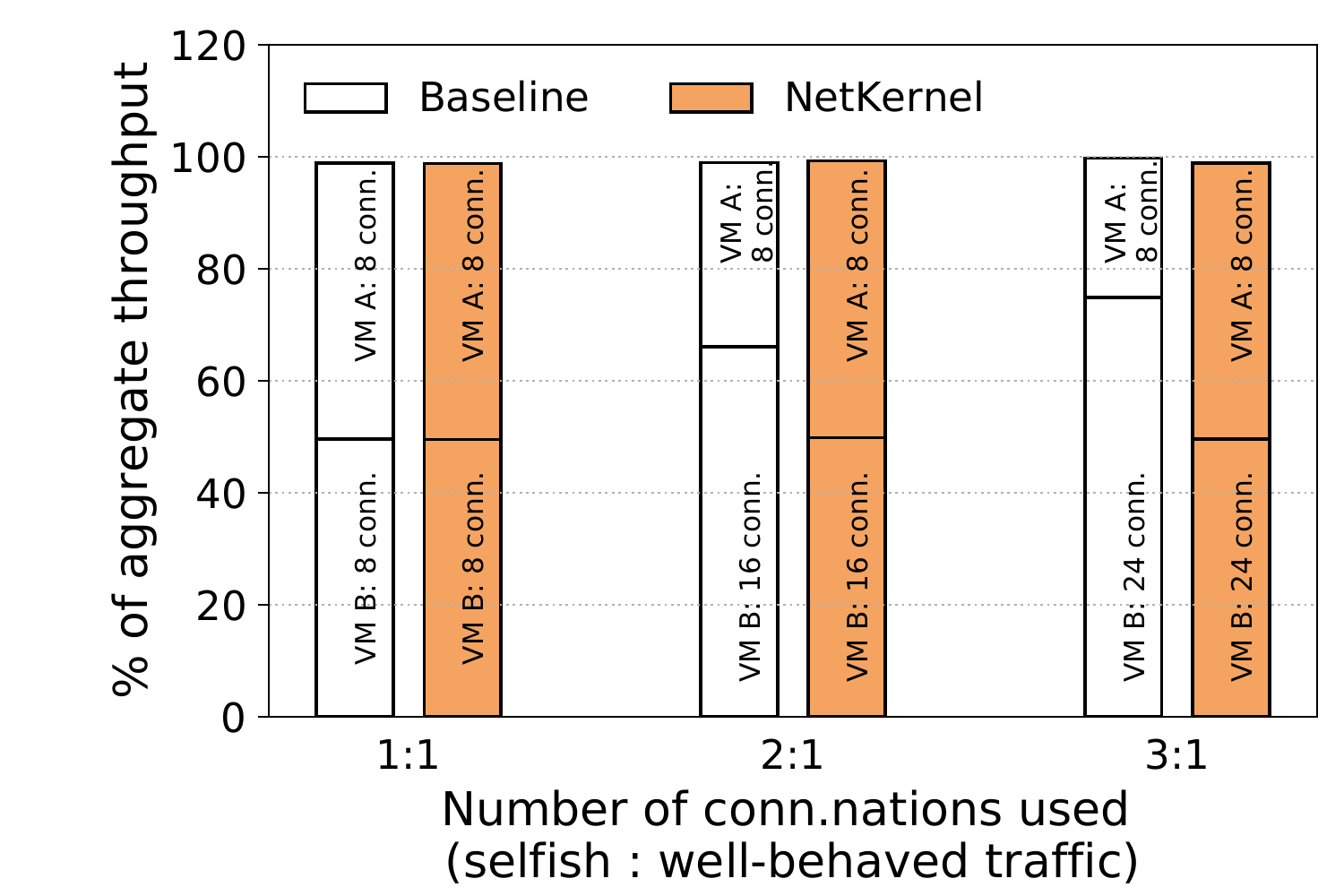}
    \caption{By sharing a unified congestion window to same destination, a NSM can achieve VM fairness.} 
    \label{fig:seawall}
\end{figure}

\sys allows schemes like Seawall to be easily implemented as
a new NSM and effectively enforce VM-level fair bandwidth sharing. Our
proof-of-concept runs a simple VM-level congestion control in the NSM: One VM
maintains a global congestion window shared among all its connections to
different destinations. Each individual flow's ACK advances the shared
congestion window, and when sending data, each flow cannot send more than
1/$n$ of the shared window where $n$ is the number of active flows. We then
run experiments with 2 VMs: a well-behaved VM that has 8 active flow, and
a selfish VM that uses varying number of active flows. Figure~\ref{fig:seawall}
presents the results. \sys with our VM-level congestion control NSM is
able to enforce an equal share of bandwidth between two VMs regardless of
number of flows. 
We leave the implementation of a complete general solution such as Seawall in
\sys as future work.

\subsection{Deploying mTCP without API Change}
\label{sec:usecase_mtcp}

We now focus on use cases of deployment and performance benefits for users.

Most userspace stacks use their own APIs and require applications to be ported \cite{JWJJ14,f-stack,Seastar:Website}. 
For example, to use mTCP an application has to use {\tt mtcp\_epoll\_wait()} to fetch events \cite{JWJJ14}. 
The semantics of these APIs are also different from socket APIs \cite{JWJJ14}.
These factors lead to expensive code changes and make it difficult to use the stack in practice.
Currently mTCP is ported for only a few applications, and does not support complex web servers like nginx. 

\begin{table}[t]
\centering
\small
\resizebox{0.65\columnwidth}{!}{
\begin{tabular}{|l|c|c|c|}
\hline
\# vCPUs & 1  & 2  & 4  \\ \hline
Kernel stack NSM  & 71.9K & 133.6K & 200.1K  \\ \hline
mTCP NSM  & 98.1K & 183.6K & 379.2K  \\ \hline
\end{tabular}
}
\caption{Performance of unmodified nginx using ab with 64B html files, a concurrency of 100, and 10M requests in total. The NSM and VM use the same number of vCPUs.}
\label{table:nginx_mtcp}
\end{table}

With \sys, applications can directly take advantage of userspace stacks without any code change. 
To show this, we deploy unmodified nginx in the VM with the mTCP NSM we implement, and benchmark its performance using ab. 
Both VM and NSM use the same number of vCPUs. 
Table~\ref{table:nginx_mtcp} depicts that mTCP provides 1.4x--1.9x improvements over the kernel stack NSM across various vCPU setting.

\sys also mitigates the maintenance efforts required from tenants. 
We provide another piece of evidence with mTCP here.
When compiling the DPDK version required by mTCP on our testbed, 
we could not set the RSS (receive side scaling) key properly 
to the mlx5\_core driver for our NIC and mTCP performance was very low.
After discussing with mTCP developers, we were able to attribute this to the asymmetric RSS key used in the NIC, 
and fixed the problem by modifying the code in DPDK mlx5 driver. 
We have submitted our fix to mTCP community. 
Without \sys tenants would have to deal with such technical complication by themselves.
Now they are taken care of transparently, saving much time and effort for many users.

\subsection{Shared Memory Networking}
\label{sec:sharedmem}
\begin{figure}[ht]
    \centering
    \includegraphics[width=0.8\linewidth]{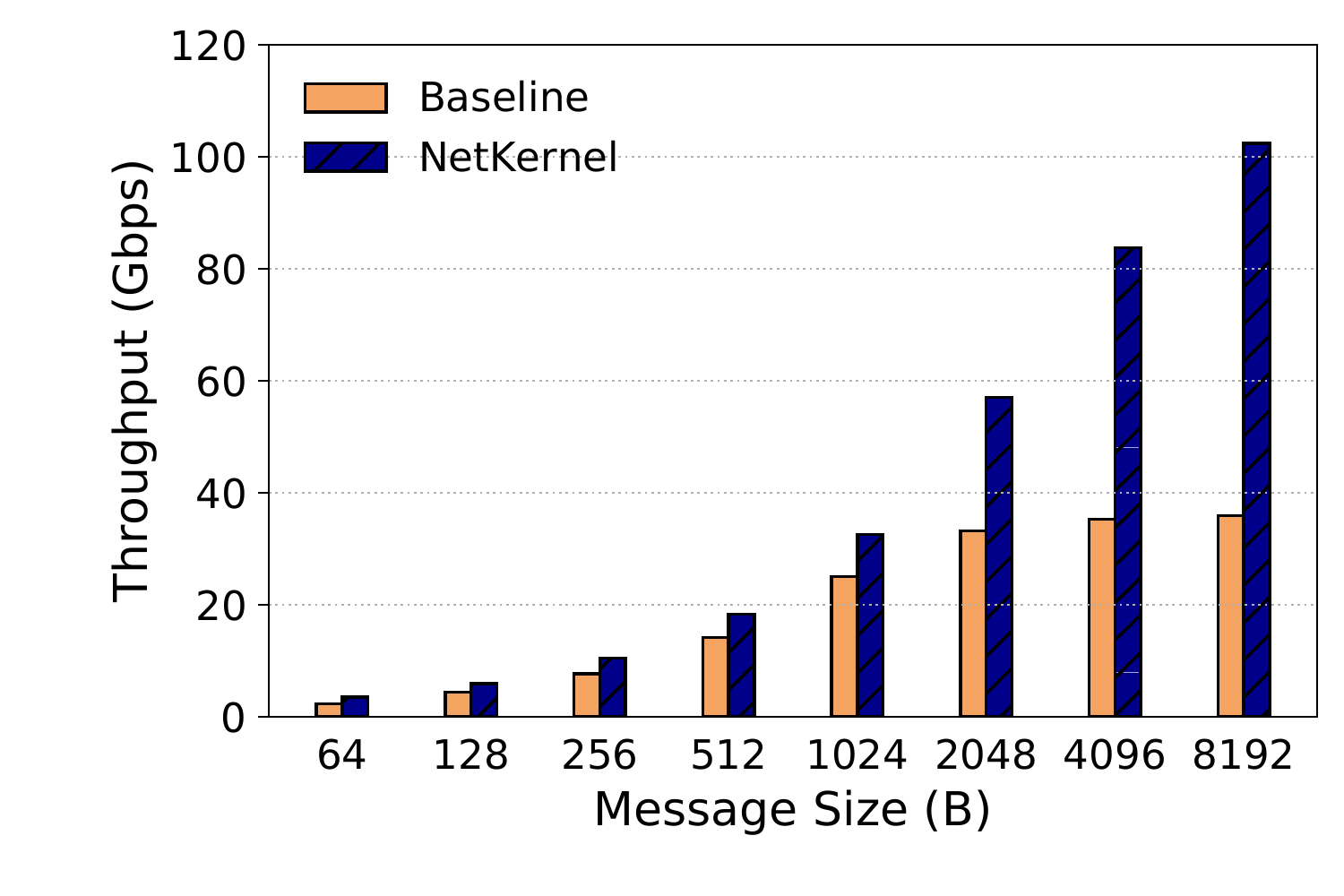}
    \caption{Using shared memory NSM for \sys for traffic between
    two colocating VMs of the same user. \sys uses 2 cores for each VM, 2 cores
    for
    the NSM, and 1 core for \controller. Baseline uses 2 core for the sending
    VM, 5 cores for receiving VM, and runs TCP Cubic. Both schemes use 8 TCP
    connections.}
    \label{fig:result_shmem}
\end{figure}

In the existing architecture, a VM's traffic always goes though its network
stack, then the vNIC, and the vSwitch, even when the other VM is on the same
host. 
It is difficult for both users and operator to optimize for this case, 
because the VM has no information about where the other endpoint is.
The hypervisor cannot help either as the data has already been processed by
the TCP/IP stack. 
With \sys the NSM is part of the infrastructure, the operator can
easily detect the on-host traffic and use shared memory to copy data for the
two VMs. 
We build a prototype NSM to demonstrate this idea: 
When a socket pair is detected as an internal socket pair by the \guestlib, and
the two VMs belong to the same user, a shared memory NSM takes over their traffic. This NSM simply copies the
message chunks between their hugepages and bypasses the
TCP stack processing. As shown in Figure~\ref{fig:result_shmem}, with 7 cores in
total, \sys with shared memory NSM can achieve 100Gbps, which is $\sim$2x
of Baseline using TCP Cubic.

%% file: evaluation.tex
\section{Evaluation}
\label{sec:evaluation}

We seek to examine a few crucial aspects of \sys in our evaluation:
(1) microbenchmarks  of \nqe switching and data copying \cref{sec:microbench}; 
(2) basic performance with the kernel stack NSM \cref{sec:basic_perf};
(3) scalability with multiple cores \cref{sec:stack_scalability} and multiple NSMs \cref{sec:nk_scalability};
(4) isolation of multiple VMs \cref{sec:isolation_perf};
(5) latency of short connections \cref{sec:latency_ms};
and (6) overhead of the system \cref{sec:overhead}.

\subsection{Setup}
\label{sec:setup}

Our testbed servers each have two Xeon E5-2698 v3 16-core CPUs clocked at 2.3~GHz, 
256~GB memory at 2133~MHz, and a Mellanox ConnectX-4 single port 100G NIC. 
Hyperthreading is disabled. 
We compare to the status quo where an application uses the kernel TCP stack in its VM, 
referred to as Baseline in the following.
We designate \sys to refer to the common setting where we use the kernel stack NSM in our implementation.
When mTCP NSM is used we explicitly mark the setting in the figures.
\controller uses one core for \nqe switching throughout the evaluation.
Unless stated otherwise, Baseline and \sys use 1 vCPU for the VM, and \sys uses 1 vCPU for the NSM.
The same TCP parameter settings are used for both systems.

\subsection{Microbenchmarks}
\label{sec:microbench}

We first microbenchmark \sys regarding \nqe and data transmission performance.

\begin{figure}[ht]
\centering
\begin{minipage}{.49\linewidth}
  \centering
  \includegraphics[width=1\linewidth]{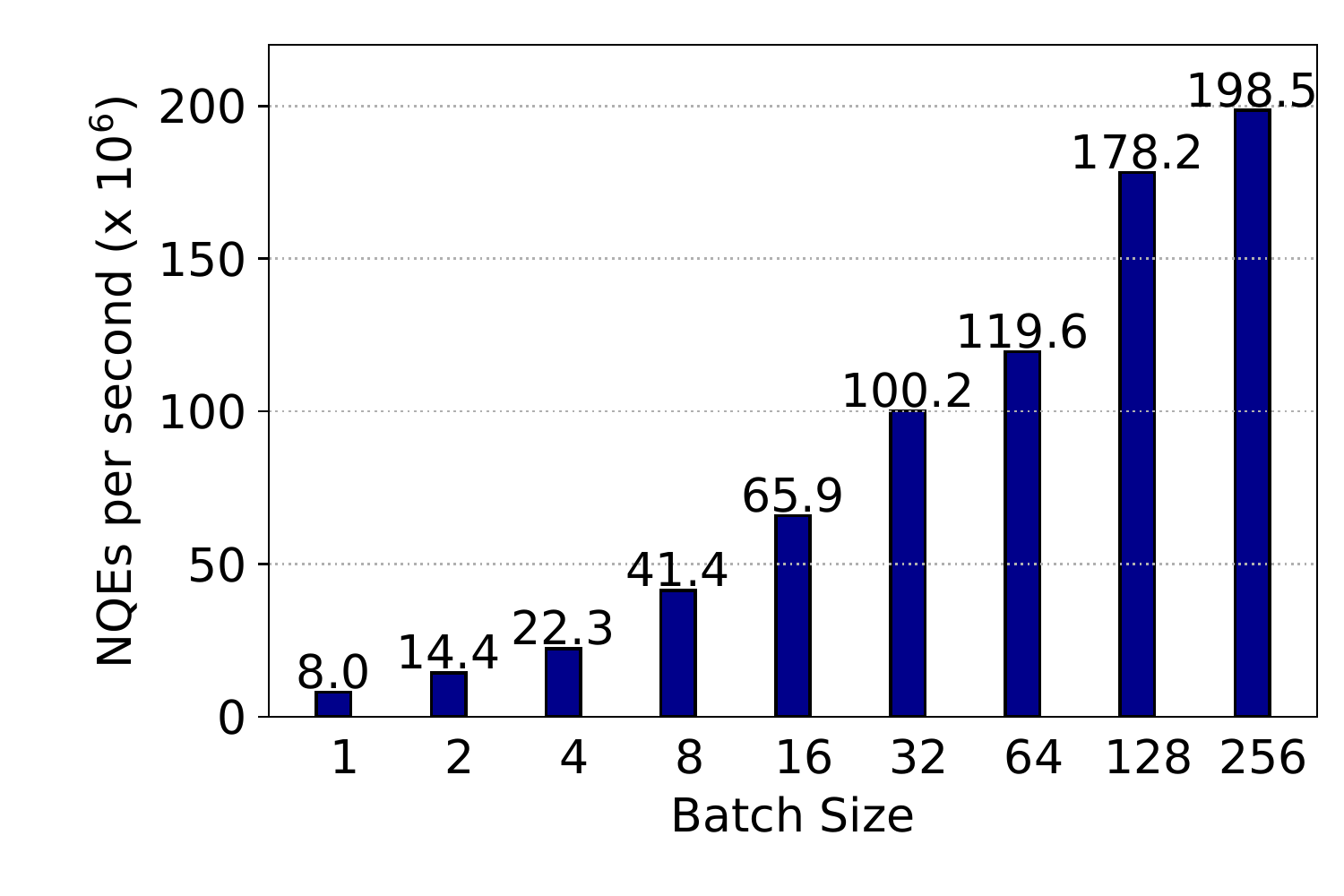}
  \caption{\controller switching throughput using a single core with different batch sizes.}
  \label{fig:result_ce_nqe}
\end{minipage} 
\begin{minipage}{.49\linewidth}
  \centering
  \includegraphics[width=1\linewidth]{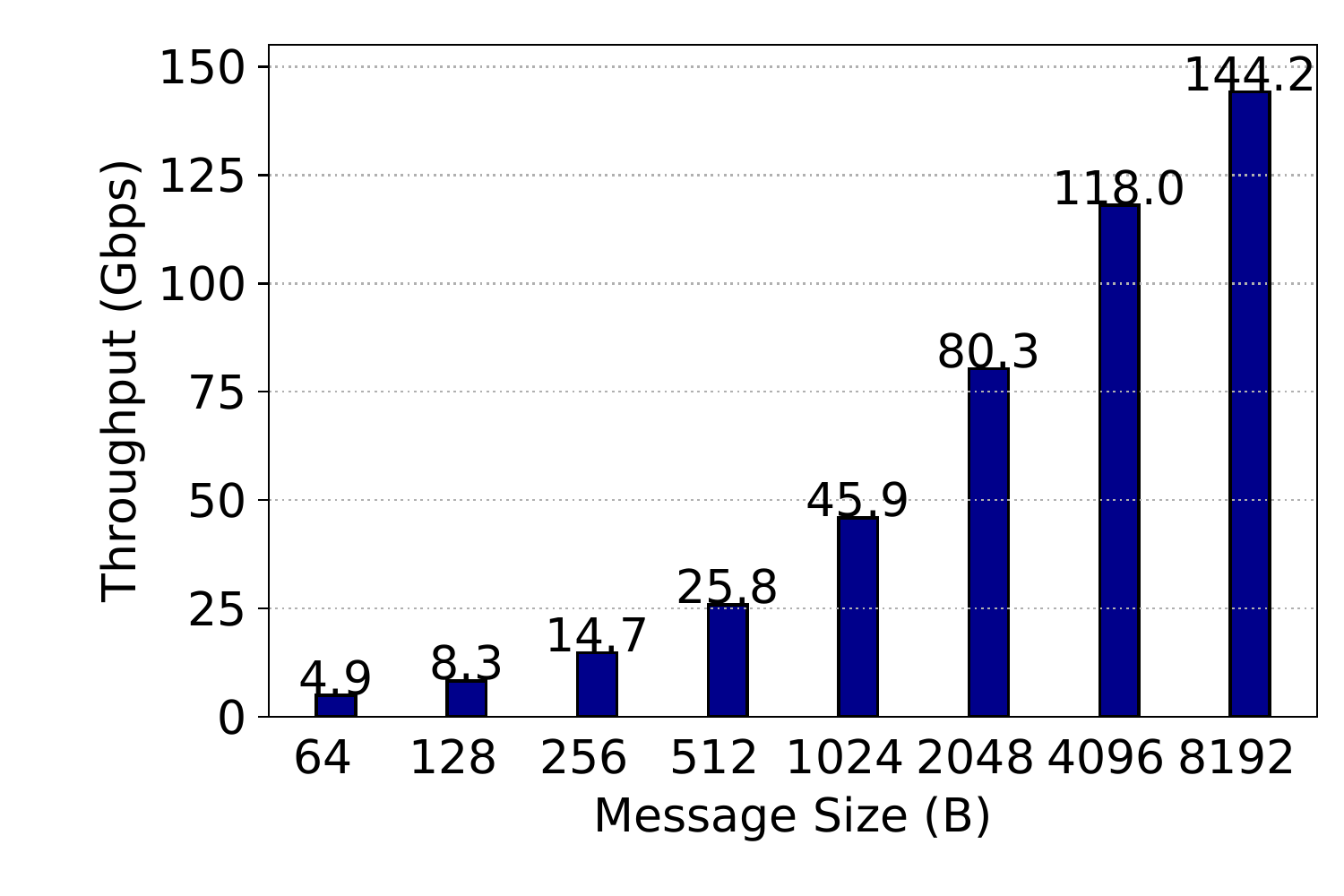}
  \captionof{figure}{Message copy throughput via hugepages with different message sizes.}
  \label{fig:result_mem_queue}
\end{minipage}
\end{figure}

\noindent{\bf \nqe switching. }
\nqes are transmitted by \controller as a software switch. It is important that \controller offers enough horsepower to ensure performance at 100G. 
We measure \controller throughput defined as the number of 32-byte \nqes copied from \guestlib's NK device queues to the \servicelib's NK device queues with two copy operations.
Figure~\ref{fig:result_ce_nqe} shows the results with varying batch sizes.
\controller achieves $\sim$8M \nqes/s throughput without batching. 
With a small batch size of 4 or 8 throughput reaches 41.4M \nqes/s and 65.9M \nqes/s, respectively, 
which is sufficient for most applications.\footnote{64Mpps provides more than 100G bandwidth with an average message size of 192B.}
More aggressive batching provides throughput up to 198M \nqes/s.
We use a batch size of 4 in all the following experiments.

\noindent{\bf Memory copy.}
We also measure the memory copy throughput between \guestlib and \servicelib via hugepages.
A memory copy in this experiment includes the following:
(0) application in the VM issues a {\tt send()} with data,
(1) \guestlib gets a pointer from the hugepages,
(2) copies the message to hugepages, 
(3) prepares a \nqe with the data pointer,
(4) \controller copies the \nqe to \servicelib,
(5) \servicelib obtains the data pointer and puts it back to the hugepages.
Thus it measures the effective application-level throughput using \sys (including \nqe transmission) without network stack processing.

Observe from Figure~\ref{fig:result_mem_queue} that \sys delivers over 100G throughput with messages larger than 4KB: with 8KB messages 144G is achievable. 
Thus \sys provides enough raw performance to the network stack and is not a bottleneck to the emerging 100G deployment in public clouds. 

\subsection{Basic Performance with Kernel Stack}
\label{sec:basic_perf}

We now look at \sys's basic performance with Linux kernel stack.
The results here are obtained with a 1-core VM and 1-core NSM; all other cores of the CPU are disabled. 
Baseline uses one core for the VM.
\begin{figure}[ht]
\centering%
\begin{minipage}{.49\linewidth}
  \centering
  \includegraphics[width=1\linewidth]{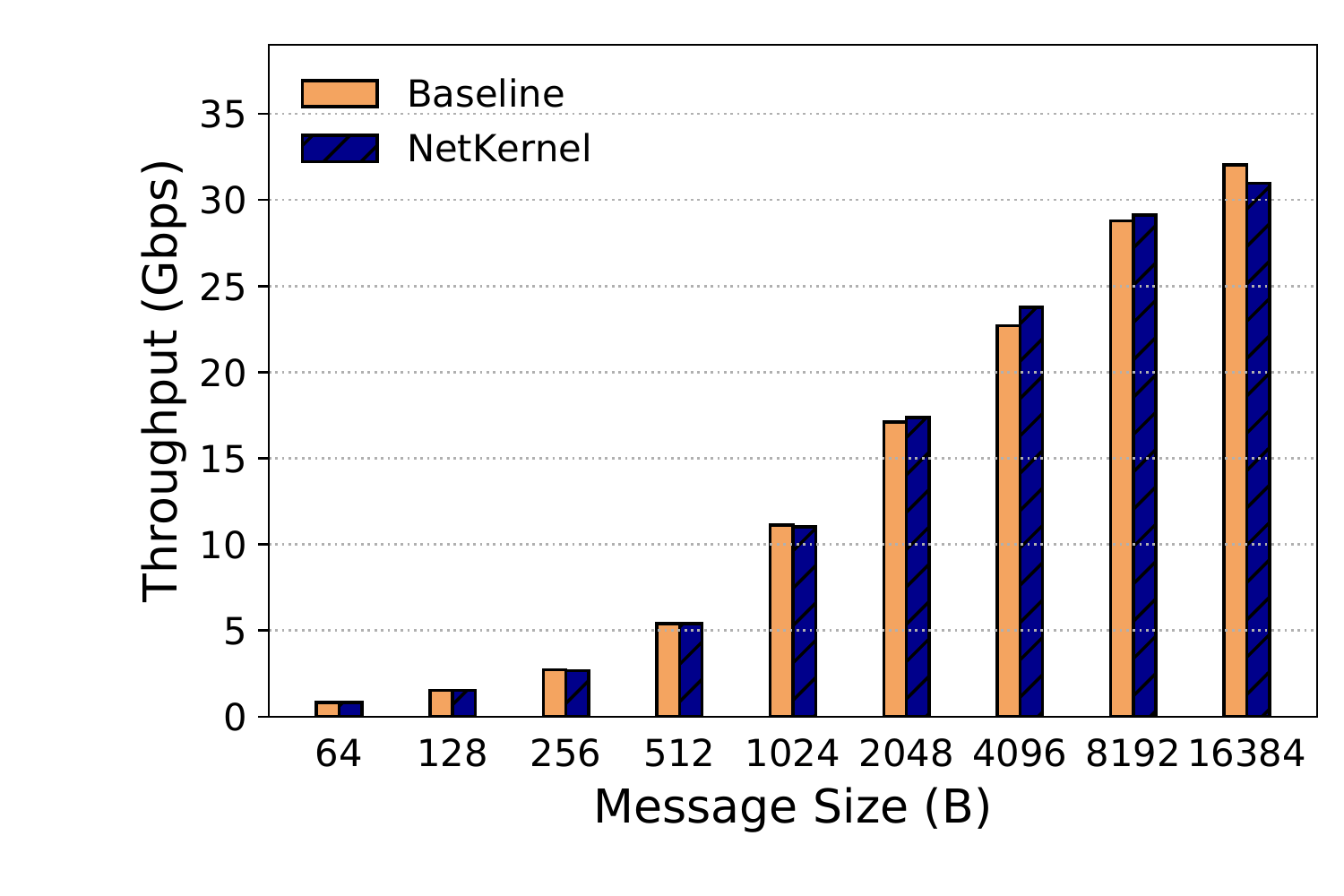}
  \caption{Single TCP stream send throughput with the kernel stack NSM. The NSM uses 1 vCPU.}
    \label{fig:result_single_tcp_send}
\end{minipage}
\begin{minipage}{.49\linewidth}
  \centering
  \includegraphics[width=1\linewidth]{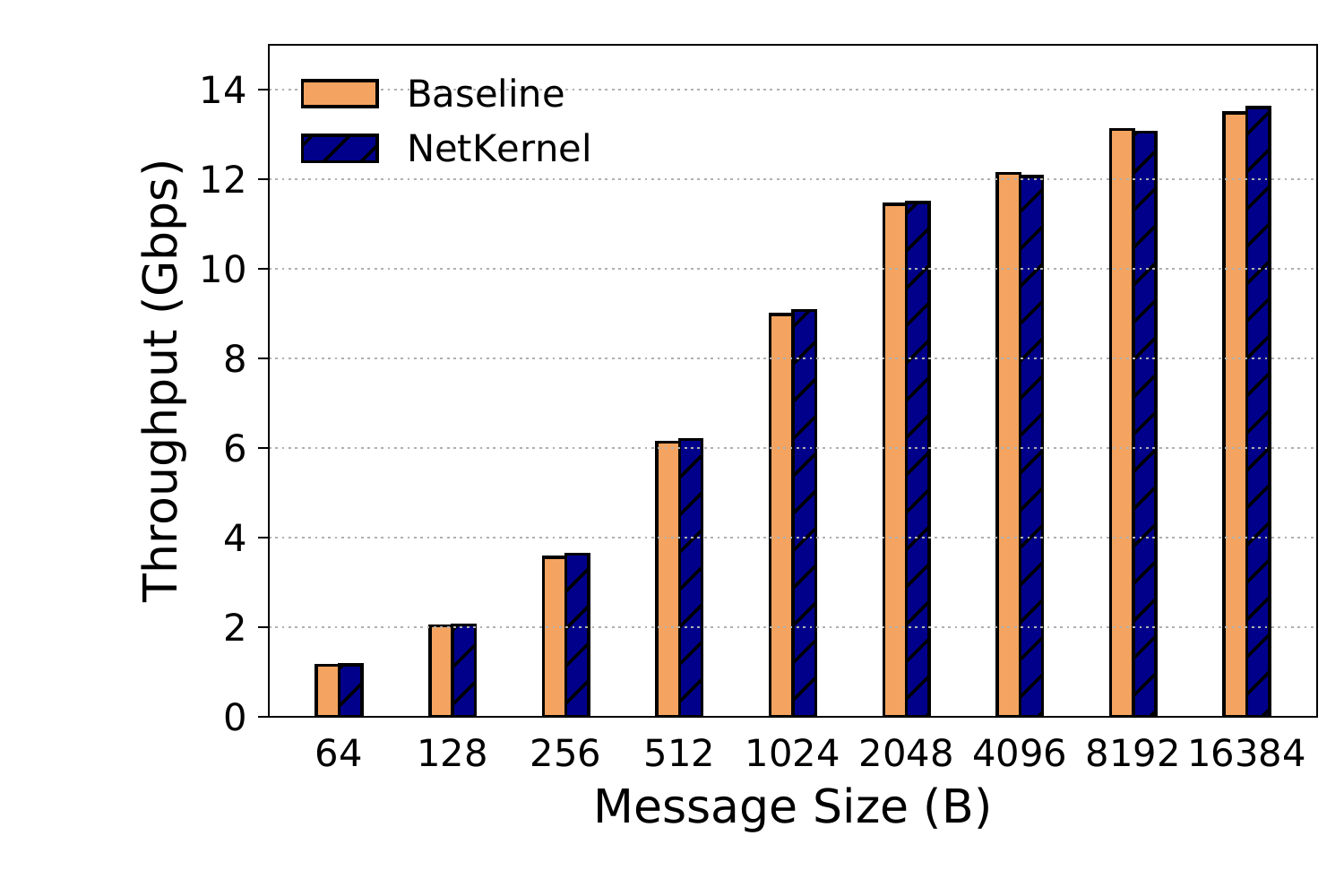}
  \caption{Single TCP stream receive throughput with the kernel stack NSM. The NSM uses 1 vCPU.}
    \label{fig:result_single_tcp_recv}
\end{minipage}
\end{figure}

\noindent{\bf Single TCP Stream.}
We benchmark the single stream TCP throughput with different message sizes. 
The results are averaged over 5 runs each lasting 30 seconds.
Figure~\ref{fig:result_single_tcp_send} depicts the send throughput and Figure~\ref{fig:result_single_tcp_recv} receive throughput. 
We find that \sys performs on par with Baseline in all cases. 
Send throughput reaches 30.9Gbps and receive throughput tops at 13.6Gbps in \sys. 
Receive throughput is much lower because the kernel stack's RX processing is much more CPU-intensive with interrupts.
Note that if the other cores of the NUMA node are not disabled, soft interrupts (softirq) may be sent to those cores instead of the one assigned to the NSM (or VM), thereby inflating the receive throughput.\footnote{We observe 30.6Gbps receive throughput with 16KB messages in both \sys and Baseline when leaving the  other cores on.}

\begin{figure}[ht]
    \centering
    \begin{minipage}{.48\linewidth}
    \includegraphics[width=\linewidth]{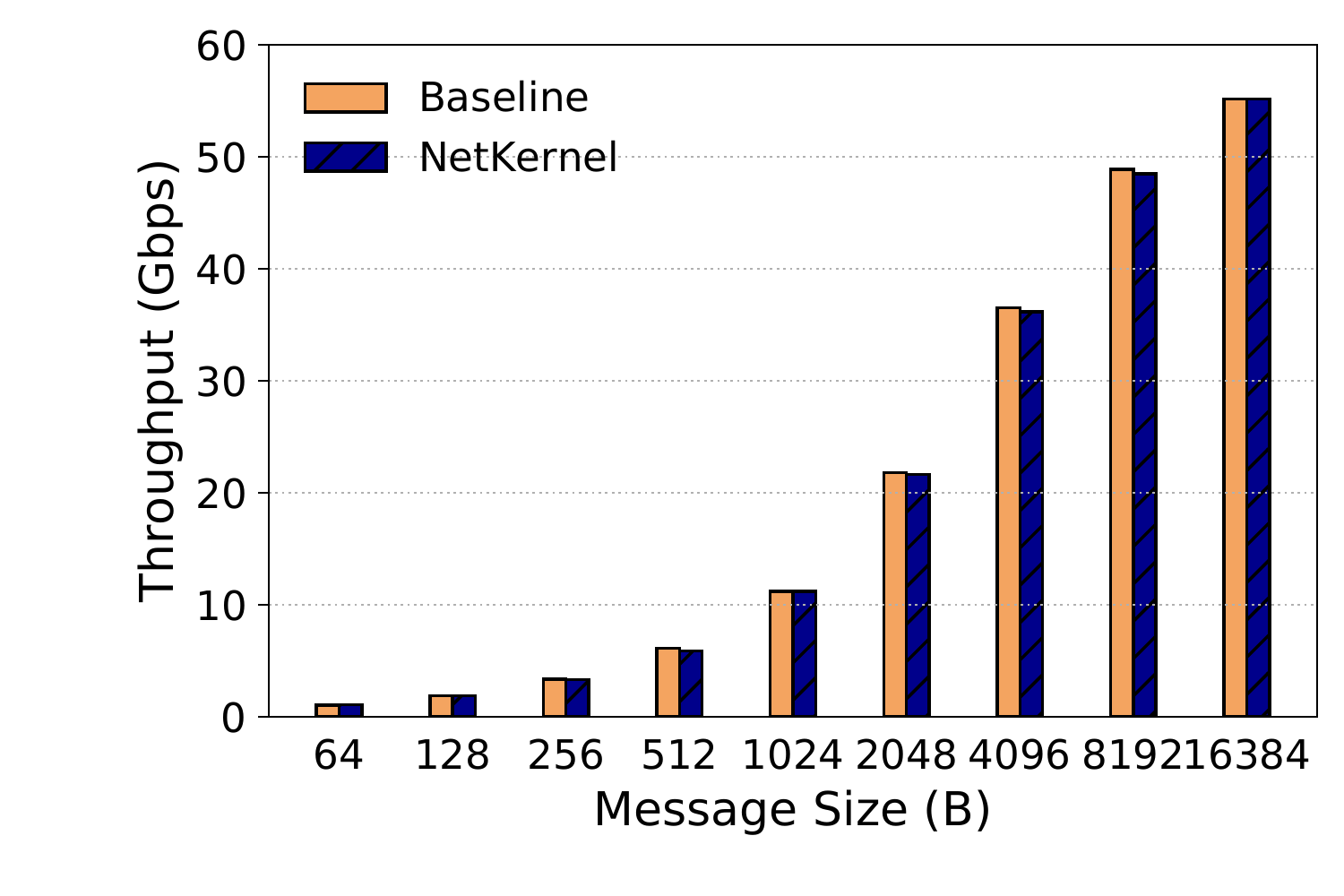}
    \caption{8-stream TCP send throughput with the kernel stack NSM. The NSM uses 1 vCPU.}
    \label{fig:result_8_tcp_send}
    \end{minipage}%
\hspace{0.1cm}
    \begin{minipage}{.48\linewidth}
    \centering
    \includegraphics[width=\linewidth]{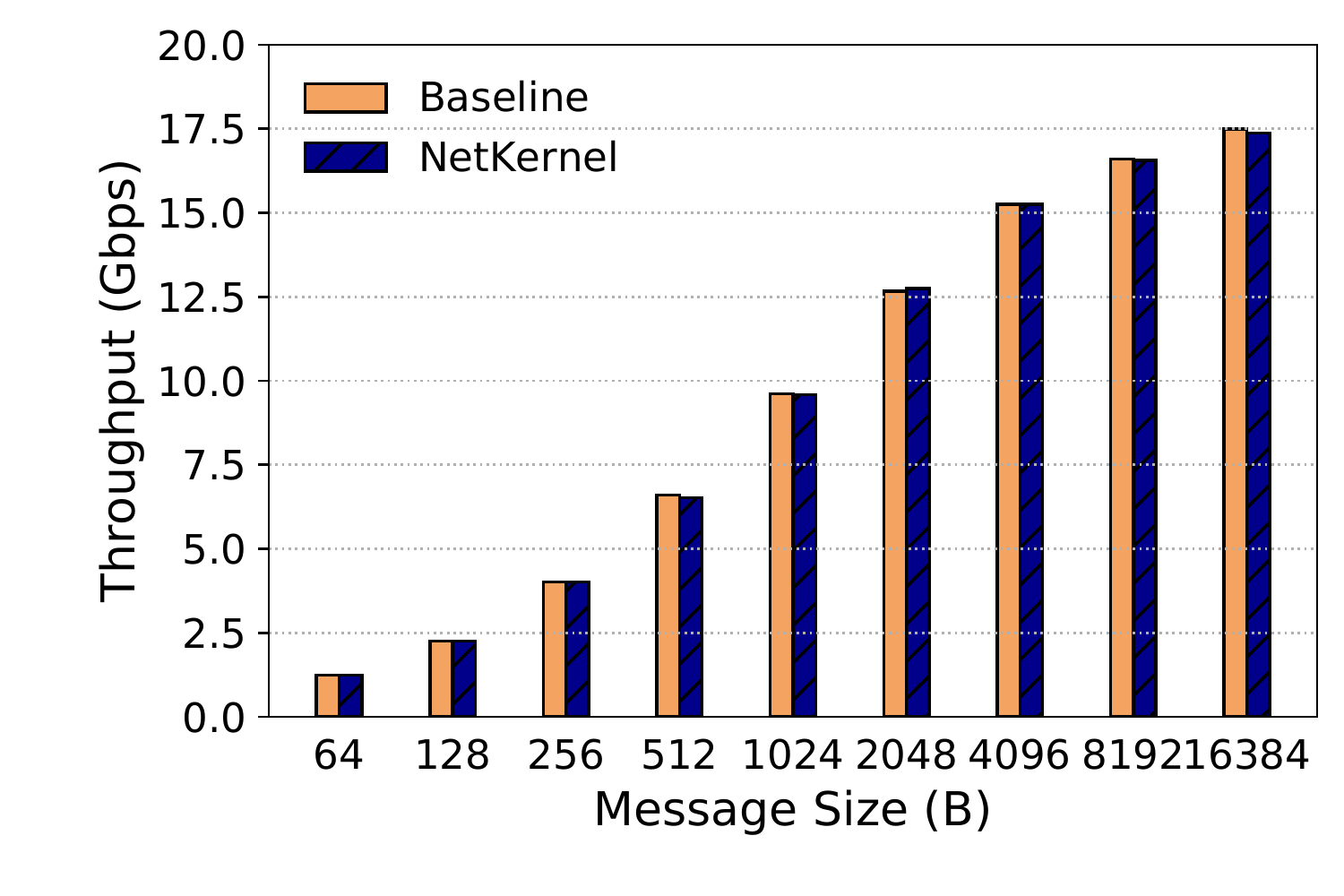}
    \caption{8-stream TCP receive throughput with the kernel stack NSM. The NSM uses 1 vCPU.}
    \label{fig:result_8_tcp_recv}
    \end{minipage}
\end{figure}

\noindent{\bf Multiple TCP Streams.}
We look at throughput for 8 TCP streams on the same single-core setup as above. 
Figures~\ref{fig:result_8_tcp_send} and \ref{fig:result_8_tcp_recv} show the results. 
Send throughput tops at 55.2Gbps, and receive throughput tops at 17.4Gbps  with 16KB messages.
\sys achieves the same performance with Baseline. 

\begin{figure}[ht]
    \centering
    \includegraphics[width=0.7\linewidth]{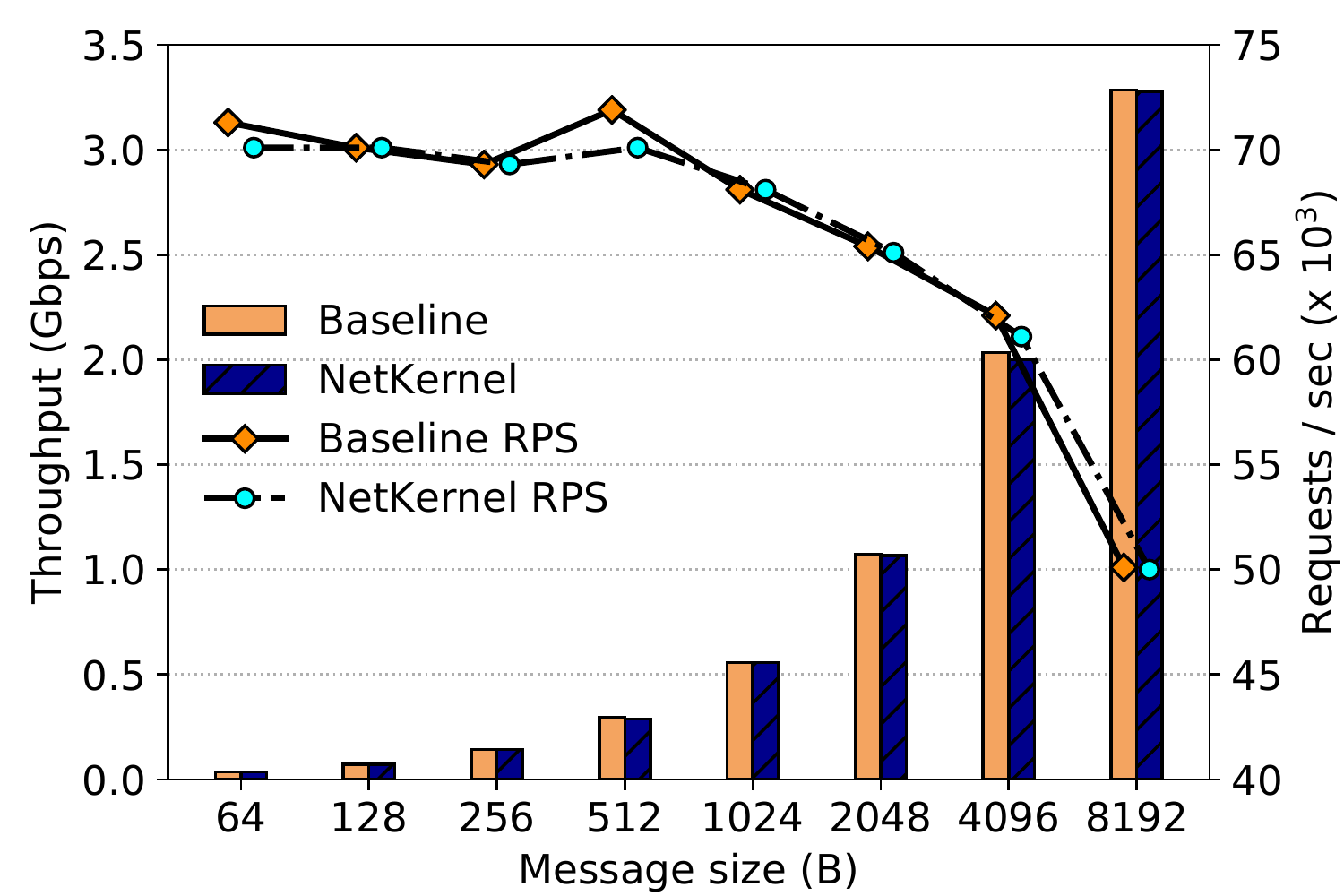}
    \caption{RPS with the kernel stack NSM using 1 vCPU. }
    \label{fig:result_rps_diff_msg_size}
    \vspace{-2mm}
\end{figure}

\noindent{\bf Short TCP Connections.}
We also benchmark \sys's performance in handling short TCP connections using a server sending a short message as a response. 
The servers are multi-threaded using epoll with a single listening port. Our workload generates 10 million requests in total with a concurrency of 1000. 
The connections are non-keepalive.
Observe from Figure~\ref{fig:result_rps_diff_msg_size} that \sys achieves $\sim$70K requests per second (rps) similar to Baseline, when the messages are smaller than 1KB. 
For larger message sizes performance degrades slightly due to more expensive memory copies for both systems.

\subsection{Network Stack Scalability}
\label{sec:stack_scalability}

Here we focus on the scalability of network stacks in \sys.

\begin{figure}[ht]
    \centering
    \begin{minipage}{.48\linewidth}
    \centering
    \includegraphics[width=\linewidth]{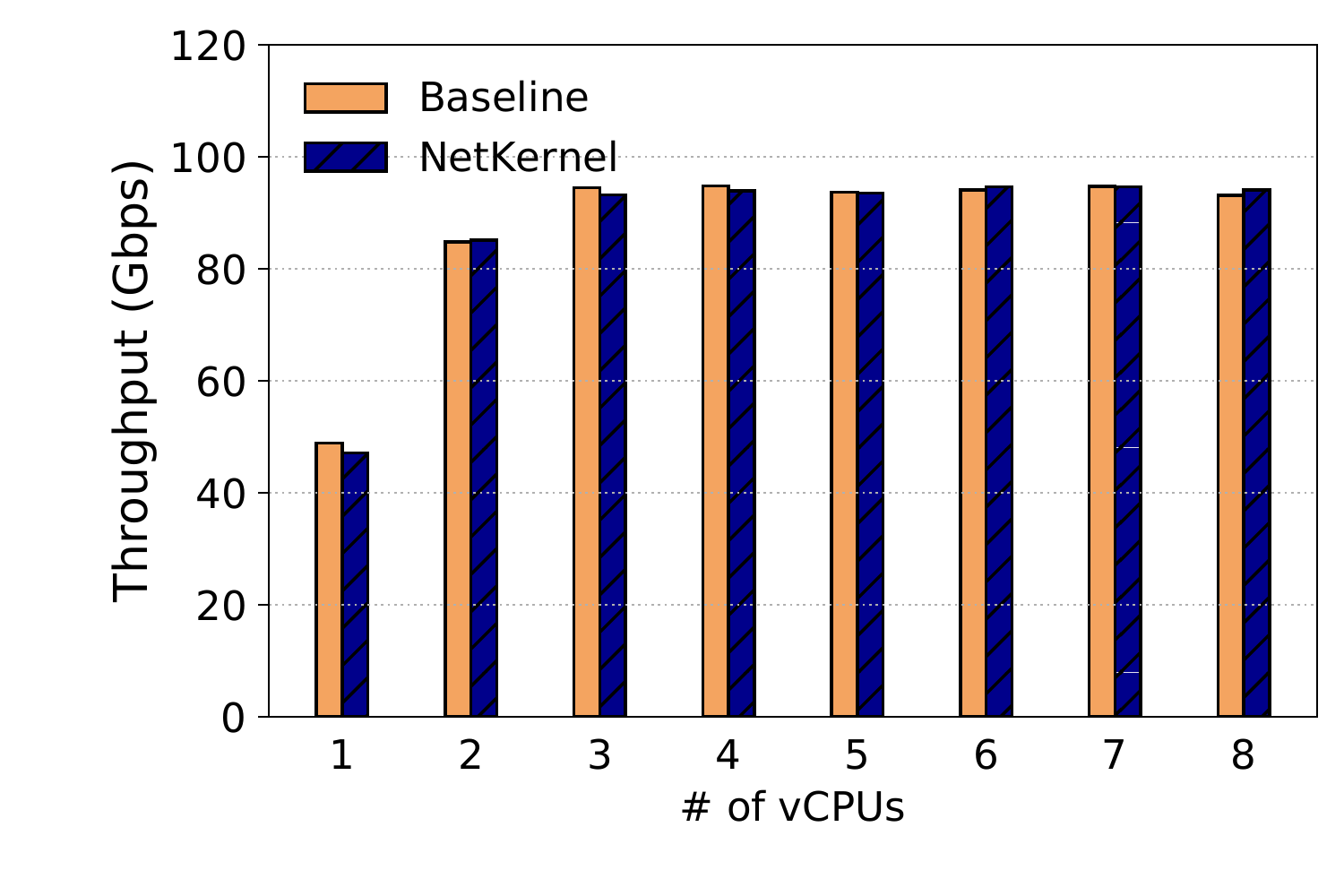}
    \caption{Send throughput of 8 TCP streams with varying numbers of vCPUs. Message size 8KB.}
    \label{fig:result_8conn_diff_noofcore_send}
    \vspace{-2mm}
    \end{minipage}
\hspace{0.1cm}
    \begin{minipage}{.48\linewidth}
    \centering
    \includegraphics[width=\linewidth]{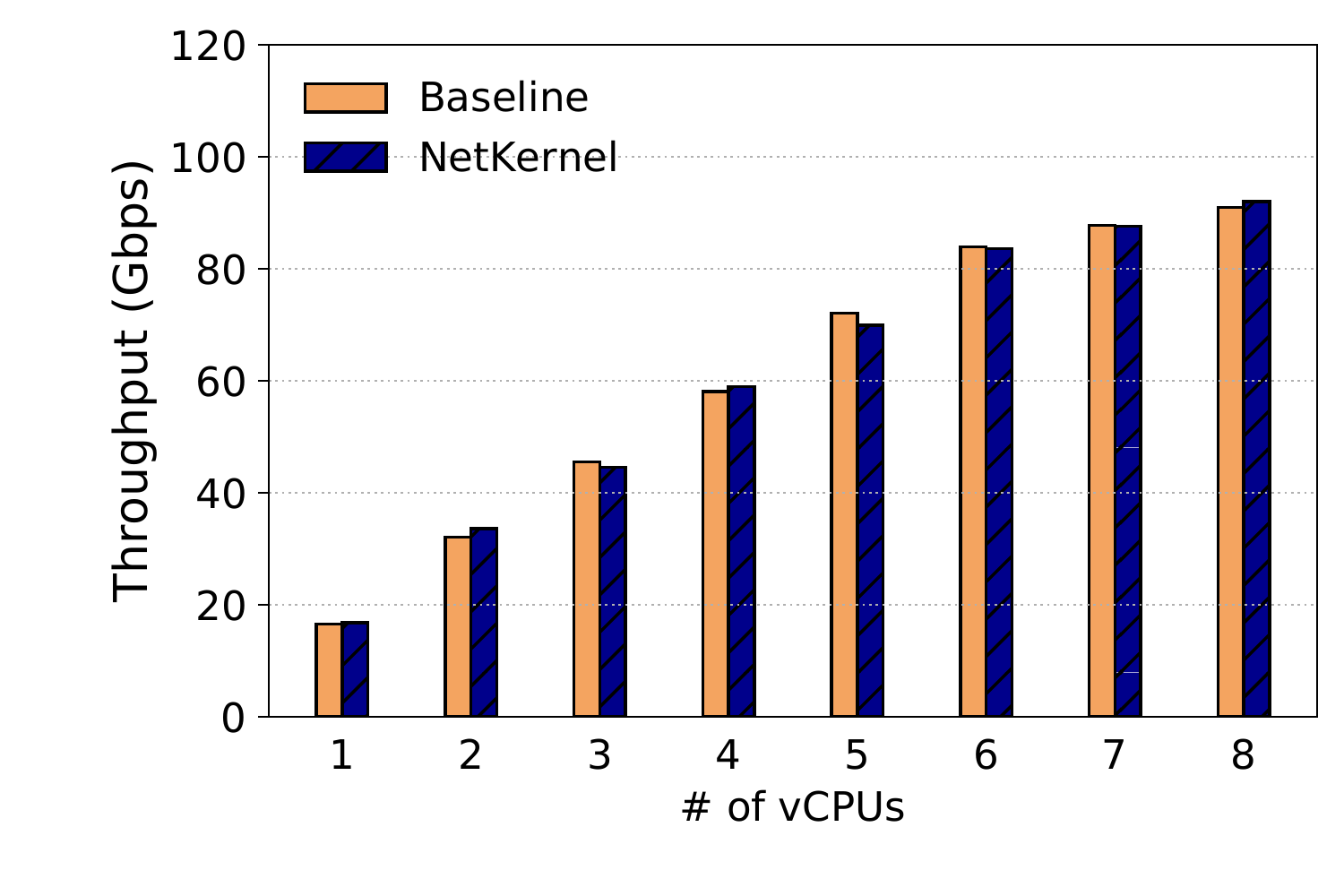}
    \caption{Receive throughput of 8 TCP streams with varying numbers of vCPUs. Message size 8KB.}
    \label{fig:result_8conn_diff_noofcore_recv}
    \vspace{-2mm}
    \end{minipage}
\end{figure}

\noindent{\bf Throughput.}
We use 8 TCP streams with 8KB messages to evaluate the throughput scalability of the kernel stack NSM. 
Results are averaged over 5 runs each lasting 30 seconds. 
Figure~\ref{fig:result_8conn_diff_noofcore_send} shows that both systems achieve the line rate of 100G using 3 vCPUs or more for send throughput. 
For receive, both achieve 91Gbps using 8 vCPUs as shown in Figure~\ref{fig:result_8conn_diff_noofcore_recv}. 

\noindent{\bf Short TCP Connections.}
We also evaluate scalability of handling short connections. 
The same epoll servers described before are used here with 64B messages.
Results are averaged over a total of 10 million requests with a concurrency of 1000.
Socket option {\tt SO\_REUSEPORT} is always used. 

\begin{figure}[ht]
    \centering
    \includegraphics[width=0.7\linewidth]{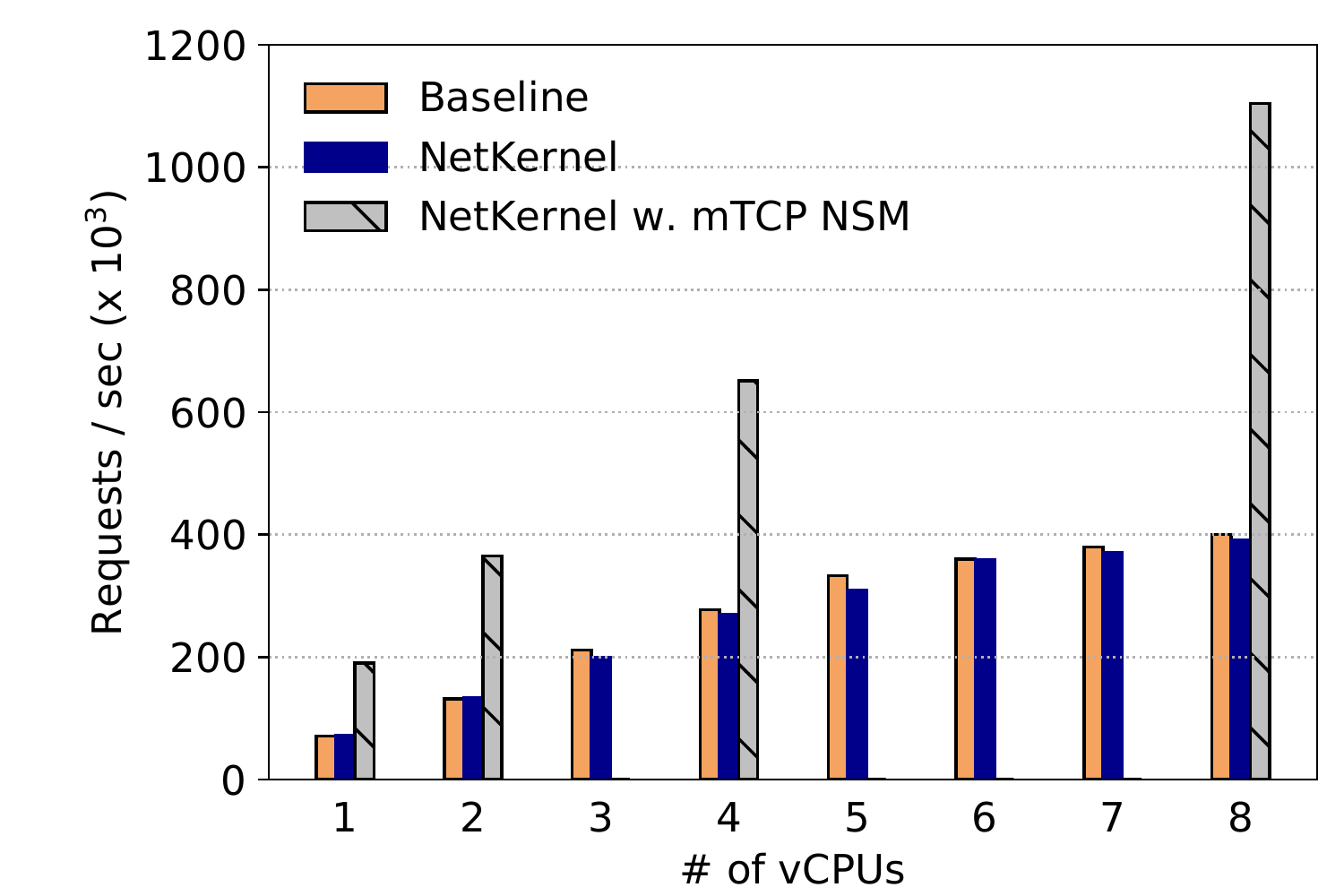}
    \caption{Performance of TCP short connections with varying number of vCPUs. Message size 64B. Both kernel stack and mTCP NSMs are used.}
    \label{fig:result_rps_epoll}
    \vspace{-2mm}
\end{figure}

Figure~\ref{fig:result_rps_epoll} shows that \sys has the same scalability as Baseline: performance increases to $\sim$400Krps with 8 vCPUs, i.e. 5.7x the single core performance. 
More interestingly, to demonstrate \sys's full capability, we also run the mTCP NSM with 1, 2, 4, and 8 vCPUs.\footnote{Using other numbers of vCPUs for mTCP causes stability problems even without \sys.}
\sys with mTCP offers 190Krps, 366Krps, 652Krps, and 1.1Mrps respectively, 
and shows better scalability than kernel stack.

The results show that \sys preserves the scalability of different network stacks, including high performance stacks like mTCP.

\subsection{\sys Scalability}
\label{sec:nk_scalability}

We now investigate the scalability of our design. 
In particular, we look at whether adding more NSMs can scale performance. 
Different from the previous section where we focus on the scalability of a network stack,
here we aim to show the scalability of \sys's overall design.

We use the same epoll servers in this set of experiments.
The methodology is the same as \cref{sec:stack_scalability}, with 8 connections and 8KB messages for throughput experiments and 10 millions of requests with 64B messages for short connections experiments.
Each kernel stack NSM now uses 2 vCPUs.
The servers in different NSMs listen on different ports and does not share an accept queue. 
We vary the number of NSMs to serve this 1-core VM.

\begin{table}[ht]
    \centering
    \small
    \resizebox{0.9\columnwidth}{!}{
    \begin{tabular}{|l|c|c|c|c|}
    \hline
    \# of 2-vCPU NSMs  & 1 & 2 & 3 & 4 \\ \hline
    Send throughput (Gbps)& 85.1 & 94.0 & 94.1 & 94.2 \\ \hline
    Receive throughput (Gbps)& 33.6 & 61.2 & 91.0 & 91.0 \\ \hline
    Requests per sec (x$10^3$) & 131.6 & 260.4 & 399.1  & 520.1 \\ \hline
    \end{tabular}
    }
    \caption{Throughput scaling and short connections with varying numbers of \sys with kernel stack NSM each with two vCPUs.}
    \label{table:result_nsm_scale}
    \vspace{4mm}
    \end{table}

Table~\ref{table:result_nsm_scale} shows the throughput scaling results. 
Throughput for send is already 85.1Gbps with 2 vCPUs (recall Figure~\ref{fig:result_8conn_diff_noofcore_send}), and adding NSMs does not improve it beyond 94.2Gbps.
Throughput for receive shows almost linear scalability on the other hand.
Performance of short connections also exhibits near linear scalability: 
One NSM provides 131.6Krps, 2 NSMs 260.4Krps, and 4 NSMs 520.1Krps which is 4x better.
The results indicate that \sys's design is highly scalable; 
reflecting on results in \cref{sec:stack_scalability}, 
the network stack's scalability limits its multicore performance.

\subsection{Isolation }
\label{sec:isolation_perf}

\begin{figure}[ht]
    \centering
    \includegraphics[width=0.7\linewidth]{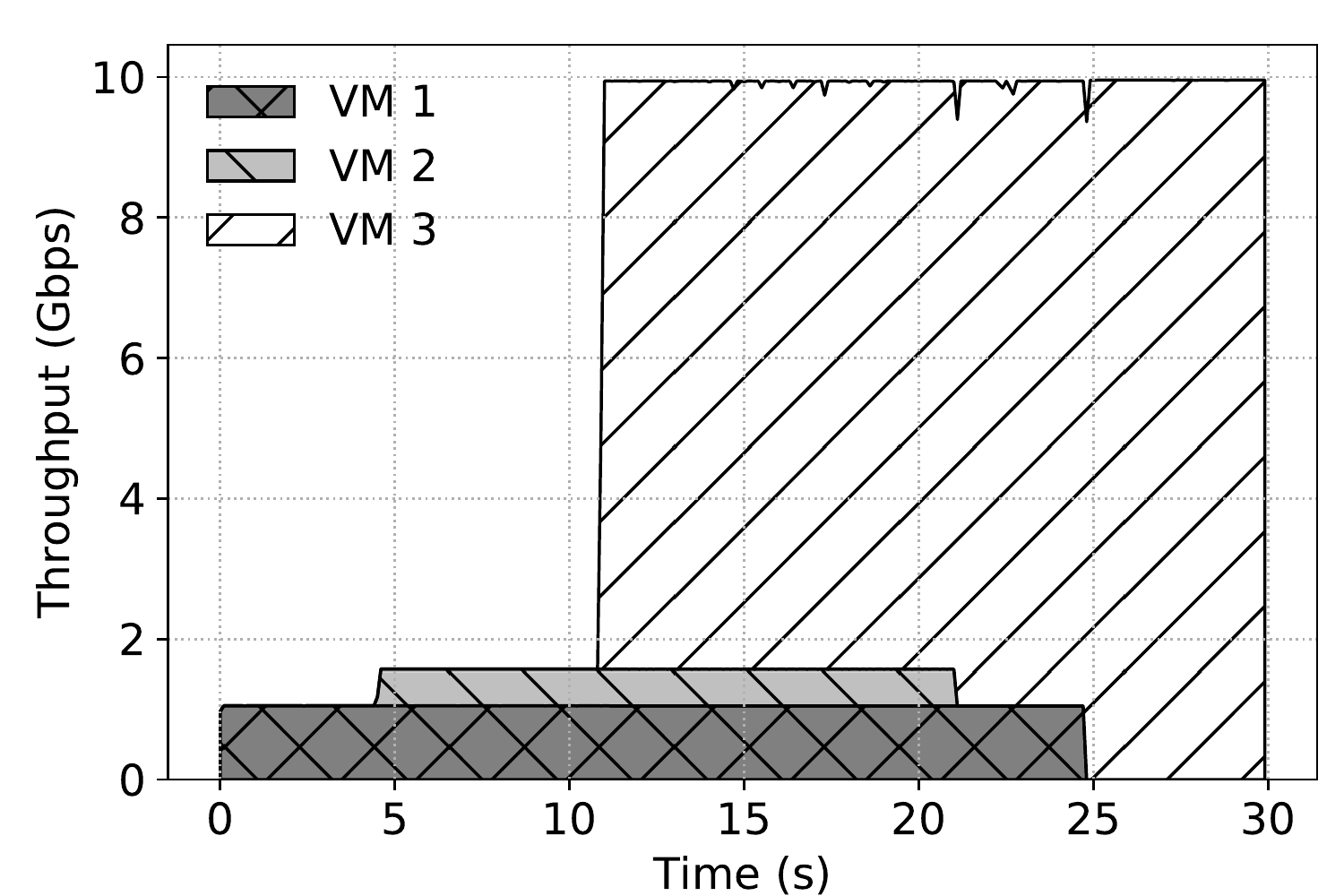}
    \caption{VM 1 is capped at 1Gbps, VM2 at 500Mbps, and VM3 uncapped. All VMs use the same kernel stack NSM. The NSM is assigned 10Gbps bandwidth. \sys isolates VM1 and VM2 successfully while allowing VM3 to obtain the remaining capacity.}
    \label{fig:result_isolation_bw}
\end{figure}

Isolation is important in public clouds to ensure co-located tenants do not interfere with each other.
We conduct an experiment to verify \sys's isolation guarantee.
As discussed in \cref{sec:ce_control}, \controller uses round-robin to poll each VM's NK device.
In addition, for this experiment we implement token buckets in \controller to limit the bandwidth of each VM, taking into account varying message sizes.
There are 3 VMs now: VM1 is rated limited at 1Gbps, VM2 at 500Mbps, and VM3 has unlimited bandwidth.
They arrive and depart at different times.
They are colocated on the same host running a kernel stack NSM using 1 vCPU.
The NSM is given a 10G VF for simplicity of showing work conservation.

Figure~\ref{fig:result_isolation_bw} shows the time series of each VM's throughput, measured by our epoll servers at 100ms intervals. 
VM1 joins the system at time 0 and leaves at 25s.
VM2 comes later at 4.5s and leaves at 21s.
VM3 joins last and stays until 30s.
Observe that \sys throttles VM1's and VM2's throughput at their respective limits correctly despite the dynamics. 
VM3 is also able to use all the remaining capacity of the 10G NSM: it obtains 9Gbps after VM2 leaves and 10Gbps after VM1 leaves at 25s. 
Therefore, \sys is able to achieve the same isolation in today's public clouds with bandwidth caps.
More complex isolation mechanisms can be applied in \sys, which is beyond the scope of this paper.

\subsection{Latency}
\label{sec:latency_ms}

One may wonder if \sys with the \nqe transmission would add delay to TCP processing, 
especially in handling short connections. 
Table~\ref{table:latency} shows the latency statistics when we run ab to generate 1K concurrent 
connections to the epoll server for 64B messages. 
A total of 5 million requests are used.
\sys achieves virtually the same latency as Baseline. 
Even for the mTCP NSM, \sys preserves its low latency due to the much simpler TCP stack processing
and various optimization \cite{JWJJ14}.
The standard deviation of mTCP latency is much smaller, implying that \sys itself provides stable 
performance to the network stacks.

\begin{table}[ht]
    \vspace{2mm}
    \centering
    \small
    \resizebox{0.9\columnwidth}{!}{
    \begin{tabular}{|l|c|c|c|c|c|}
    \hline
     & Min & Mean & Stddev & Median & Max\\ \hline
    Baseline  & 0 & 16 & 105.6 & 2 & 7019\\ \hline
    \sys  & 0 & 16 & 105.9 & 2 & 7019 \\ \hline
    \sys, mTCP NSM & 3 & 4 & 0.23 & 4 & 11 \\ \hline
    \end{tabular}
    }
    \caption{Distribution of response times (ms) for 64B messages with 5 million requests and 1K concurrency.}
    \label{table:latency}
    \vspace{2mm}
    \end{table}

\subsection{Overhead}
\label{sec:overhead}

We finally investigate \sys's CPU overhead. 
To quantify it, we use the epoll servers at the VM side, 
and run clients from a different machine with fixed throughput or requests per second for both \sys and Baseline with kernel TCP stack.
We disable all unnecessary background system services in both the VM and NSM, and ensures the CPU usage is almost zero without running our servers.
During the experiments, we measure the total number of cycles spent by the VM in Baseline,
and the total cycles spent by the VM and the NSM together in \sys. 
We then report \sys's CPU usage normalized over Baseline's for the same performance level in Tables~\ref{table:overhead_throughput} and \ref{table:overhead_rps}.

\begin{table}[ht]
\centering
\small
\resizebox{\columnwidth}{!}{
\begin{tabular}{|l|c|c|c|c|c|}
\hline
Throughput & 20Gbps & 40Gbps & 60Gbps & 80Gbps & 100Gbps \\ \hline
Normalized CPU usage  & 1.14 & 1.28 & 1.42  & 1.56 & 1.70 \\ \hline
\end{tabular}
}
\caption{Overhead for throughput. The NSM runs the Linux kernel TCP stack. We use 8 TCP streams with 8KB messages. \sys's CPU usage is normalized over that of Baseline.}
\label{table:overhead_throughput}
 \vspace{-2mm}
\end{table}

\begin{table}[ht]
\centering
\small
\resizebox{0.9\columnwidth}{!}{
\begin{tabular}{|l|c|c|c|c|c|}
\hline
Requests per second (rps) & 100K  & 200K  & 300K  & 400K  & 500K  \\ \hline
Normalized CPU usage  & 1.06 & 1.05 & 1.08  & 1.08 & 1.09 \\ \hline
\end{tabular}
}
\caption{Overhead for short TCP connections. The NSM runs the Linux kernel TCP stack. We use 64B messages with a concurrency of 100. }
\label{table:overhead_rps}
\end{table}

We can see that to achieve the same throughput, \sys incurs relatively high overhead especially
as throughput increases. 
This is due to the extra memory copy from the hugepages to the NSM.  
This overhead can be optimized away by implementing zerocopy between the hugepages and the NSM, 
which we are working on currently.

Table~\ref{table:overhead_rps} shows \sys's overhead with short TCP connections.
Observe that the overhead ranges from 5\% to 9\% in all cases and is fairly mild.
As the message is only 64B here, the results verify that the \nqe transmission overhead 
of the NK devices is small. 

Lastly, throughout all experiments of our evaluation we dedicated one core to \controller, which is another overhead.
As we focus on showing  feasibility and potential of  \sys in this paper, we resort to software \nqe switching
which attributes to the polling overhead. 
It is possible to explore hardware offloading using FPGAs for example to attack this overhead, just like offloading the vSwitch processing to SmartNICs \cite{FPMC18,F17}. 
This way \controller does not consume CPU for the majority of the \nqes: 
only the first \nqe of a new connection needs to be handled in CPU (direct to a proper NSM as in \cref{sec:nqe_switching}). 

To quickly recap, current \sys implementation incurs CPU overheads especially for extra data copy and \controller.
We believe, however, they are not inevitable when separating the network stack from the guest OS.
They can be largely mitigated using known implementation techniques which is left as future work.  
In addition, as shown in \cref{sec:multiplex} multiplexing can be used in \sys's current implementation to actually save CPU compared to dedicating cores to individual VMs. 

%% file: discussion.tex

\section{Discussion}
\label{sec:discussion}



\sys marks a significant departure from the way networking is provided to VMs nowadays. 
One may have the following concerns which we address now. 


\noindent{\bf How about security?}
One may have security concerns with \sys's approach of using the
provider's NSM to handle tenant traffic. Security impact is minimal because
most of the security protocols such as HTTPS/TLS work at the application
layer. They can work as usual with \sys. One exception is IPSec. Due to
the certificate exchange issue, IPSec does not work directly in our design.
However, in practice IPSec is usually implemented at dedicated gateways
instead of end-hosts \cite{SXTW17}. Thus we believe the impact is not serious.

\noindent{\bf How about fate-sharing?}
Making network stack a service introduces some more additional fate-sharing,
say when VMs share the same NSM. We believe this is not serious because cloud
customers already have fate-sharing with the vSwitch, hypervisor, and the
complete virtual infrastructure. The efficiency, performance, and convenience
benefits of our approach as demonstrated before outweigh the marginal increase
of fate-sharing; the success of cloud computing these years is another strong
testament to this tradeoff.

\noindent{\bf How can I do netfilter now?}
Due to the removal of vNIC and redirection from the VM's own TCP stack, some
networking tools like netfilter are affected. Though our current design does
not address them, they may be supported by adding additional callback
functions to the network stack in the NSM. When the NSM serves multiple VMs,
it then becomes challenging to apply netfilter just for packets of a specific
VM. We argue that this is acceptable since most tenants wish to focus on their
applications instead of tuning a network stack. \sys does not aim to
completely replace the current architecture. Tenants may still use the VMs
without \sys if they wish to gain maximum flexibility on the network
stack implementation.




\noindent{\bf Can hardware offloading be supported?}
Providers are exploring how to offload certain networking tasks, such as
congestion control, to hardware like FPGA \cite{AGRW17} or programmable NICs
\cite{NCGN17}. \sys is not at odds with this trend. It actually provides
better support for hardware offloading compared to the legacy architecture.
The provider can fully control how the NSM utilizes the underlying hardware
capabilities. \sys can also exploit hardware acceleration for \nqe
switching as discussed in \cref{sec:overhead}.

%% file: related.tex

\section{Related Work}
\label{sec:design_spectrum}

We survey several lines of closely related work.

There has been emerging interest on providing proper congestion control
abstractions in our community. CCP \cite{NCRG18} for examples puts forth a
common API to expose various congestion control signals to congestion control
algorithms independent of the data path. HotCocoa proposes abstractions for
offloading congestion control to hardware \cite{AGRW17}. They focus on
congestion control while \sys focuses on  stack architecture. They are thus
orthogonal to our work and can be deployed as NSMs in \sys to reduce the
effort of porting different congestion control algorithms.

Some work has looked at how to enforce a uniform congestion control logic across tenants without modifying VMs \cite{he2016ac,CBVR16}. The differences between this line of work and ours are clear: these approaches require packets to go through two different stacks, one in the guest kernel and another in the hypervisor, leading to performance and efficiency loss. \sys does not suffer from these problems. 
In addition, they also focus on congestion control while our work targets the entire network stack. 

In a broader sense, our work is also related to the debate on how an OS should be architected in general, and microkernels \cite{golub1992microkernel} and unikernels \cite{EKO95,madhavapeddy2013unikernels} in particular.
Microkernels take a minimalist approach and only implement address space management, thread management, and IPC in the kernel. Other tasks such as file systems and I/O are done in userspace \cite{VTZ17}. 
Unikernels \cite{madhavapeddy2013unikernels,EKO95} aim to provide various OS services as libraries that can be flexibly combined to construct an OS. 
Different from these works that require radical changes to the OS, we seek to flexibly provide the network stack as a service without re-writing the existing guest kernel or the hypervisor. 
In other words, our approach brings some key benefits of microkernels and unikernels without a complete overhaul of existing virtualization technology.
Our work is also in line with the vision presented in the position paper \cite{NXHC17}. 
We provide the complete design, implementation, and evaluation of a working system in addition to several new use cases compared to  \cite{NXHC17}.

Lastly, there are many novel network stack designs that improve performance. 
The kernel TCP/IP stack continues to witness optimization efforts in various aspects \cite{YHSE16,pathak2015modnet,lin2016scalable}. 
On the other hand, since mTCP \cite{JWJJ14} userspace stacks based on high performance packet I/O have been quickly gaining momentum \cite{Seastar:Website,marinos2014network,PLZP14,ZEFG15,MLDB15,OpenOnload,YNRL16}. 
Beyond transport layer, novel flow scheduling \cite{BCCH15} and end-host based load balancing schemes \cite{HRAF15,KHGK16} are developed to reduce flow completion times.
These proposals are for specific problems of the stack, and can be potentially deployed as network stack modules in \sys. 
This paper takes on a broader and fundamental issue: how can we properly re-factor the VM network stack, so that different designs can be easily deployed, and operating them can be more efficient?